\RequirePackage{lineno}
\documentclass[journal]{IEEEtran}
\usepackage{threeparttable}
\usepackage{lipsum}
\usepackage[dvips]{color}
\usepackage{graphicx}
\usepackage{lipsum}
\usepackage{algorithm,algorithmic}
\usepackage{epstopdf}
\usepackage{amsmath,amsfonts,cite}
\usepackage{booktabs}
\newcommand{\thickhline}{\noalign{\hrule height 1.0pt}}
\usepackage{multirow}
\ifCLASSINFOpdf
 
\else
\fi
\begin{document}
\title{Stochastic Testing Method for Transistor-Level Uncertainty Quantification Based on Generalized Polynomial Chaos}

\author{Zheng Zhang, Tarek A. El-Moselhy, Ibrahim (Abe) M. Elfadel, and~Luca~Daniel

\thanks{This work was supported by the Cooperative Agreement Between the Masdar Institute of Science and Technology, Abu Dhabi, UAE and the Massachusetts Institute of Technology (MIT), Cambridge, MA, USA, Reference No.196F/002/707/102f/70/9374.}
\thanks{Z. Zhang and L. Daniel are with the Department of Electrical Engineering and Computer Science, Massachusetts Institute of Technology (MIT), Cambridge, MA 02139, USA. E-mail: z\_zhang@mit.edu, luca@mit.edu.}
\thanks{T. El-Moselhy is with the Department of Aeronautics and Astronautics, MIT, Cambridge, MA 02139, USA. E-mail: tmoselhy@mit.edu}
\thanks{I. M. Elfadel is with Masdar Institute of Science and Technology, Abu Dhabi, United Arab Emirates. E-mail: ielfadel@masdar.ac.ae.}
}

\markboth{IEEE TRANSACTIONS ON COMPUTER-AIDED DESIGN OF INTEGRATED CIRCUITS AND SYSTEMS, ~Vol. ~XX, No.~XX,~XX~2013}{ZHANG \MakeLowercase{\textit{et al.}}: ST Method for Transistor-Level UQ based on gPC}

\maketitle

\begin{abstract}
Uncertainties have become a major concern in integrated circuit design. In order to avoid the huge number of repeated simulations in conventional Monte Carlo flows, this paper presents an intrusive spectral simulator for statistical circuit analysis. Our simulator employs the recently developed generalized polynomial chaos expansion to perform uncertainty quantification of nonlinear transistor circuits with both Gaussian and non-Gaussian random parameters. We modify the nonintrusive stochastic collocation (SC) method and develop an intrusive variant called stochastic testing (ST) method to accelerate the numerical simulation. Compared with the stochastic Galerkin (SG) method, the resulting coupled deterministic equations from our proposed ST method can be solved in a decoupled manner at each time point. At the same time, ST uses fewer samples and allows more flexible time step size controls than directly using a nonintrusive SC solver. These two properties make ST more efficient than SG and than existing SC methods, and more suitable for time-domain circuit simulation. Simulation results of several digital, analog and RF circuits are reported. Since our algorithm is based on generic mathematical models, the proposed ST algorithm can be applied to many other engineering problems.
\end{abstract}

\IEEEpeerreviewmaketitle
Ä
\section{Introduction}
\IEEEPARstart{V}{ariation} has become a major concern in today's nanometer integrated circuit design~\cite{variation2008}. It is well known that the uncertainties of transistor threshold voltages have significantly limited the scaling down of the supply voltage in low-power design~\cite{jssc1995,VLSI2008}. Meanwhile, manufacturing uncertainties can remarkably influence the performance of on-chip interconnects~\cite{Tarek_DAC:08,Tarek_DAC:10,Tarek_ISQED:11,TarekPhD,cmpt2012,igor:2012,zzhang_iccad:2011,LDaniel:2004}, leading to timing variations~\cite{IEDM98,DAC04timing}. These device-level uncertainties can propagate to the circuit or system level, and finally influence chip performance and yield~\cite{yield1995}. Therefore, new electronic design automation (EDA) tools are highly desirable to model and simulate the uncertainties at different levels~\cite{cicc2011,Boning00modelsof,Tarek_DAC:08,Tarek_DAC:10,Tarek_ISQED:11,TarekPhD,cmpt2012,igor:2012,xli2010,TCAD2006,wyzhang2011}.

One bottleneck lies in quantifying the uncertainty propagating from the device level to the circuit or system level. Such uncertainty quantification (UQ) problems require specialized stochastic solvers to estimate the underlying statistical information by detailed transistor-level simulation. The mainstream transistor-level simulators such as PSpice~\cite{Tuinenga:1995}, Cadence Spectre~\cite{Cadence}, and Synopsys HSPICE~\cite{HSPICE} utilize the well-known Monte Carlo (MC) algorithm~\cite{MonteCarlo} to perform a statistical characterization. Unfortunately, MC must run repeated transistor-level simulations at a huge number of sampling points due to its slow convergence rate. Although some improvements have been proposed (such as Quasi-Monte Carlo and Latin Hypercube samplings~\cite{Kanj:2006,SingheeR09,SingheeR10}), MC simulation is still inefficient for many circuit UQ problems.

As an alternative, spectral methods based on polynomial chaos (PC) expansions have been proposed to accelerate the UQ of circuits with Gaussian random parameters~\cite{Tarek_DAC:08,Tarek_DAC:10,Tarek_ISQED:11,TarekPhD,cmpt2012,igor:2012,Tao:2007,Strunz:2008}. Spectral methods represent the circuit uncertainties by truncated Hermite-chaos polynomial~\cite{PC1938} [which is abbreviated to polynomial chaos (PC)] series expansions, and they compute the PC coefficients by a stochastic Galerkin (SG)~\cite{sfem} or stochastic collocation (SC)~\cite{col:2005} approach. The intrusive SG method solves a coupled deterministic equation by modifying an existing deterministic solver to directly compute the PC coefficients. Alternatively, the nonintrusive SC scheme solves a set of decoupled equations at some sampling points by repeatedly calling an existing deterministic solver, followed by a numerical procedure to reconstruct the PC coefficients. Since the truncated PC expansion converges very fast when the solution dependence on the random parameters is smooth, spectral methods have shown remarkable speedup over MC when the number of parameters is small or medium. In the context of EDA, most work has been focused on applying SC and SG to solve the linear stochastic equations arising from interconnect analysis~\cite{Tarek_DAC:08,Tarek_DAC:10,Tarek_ISQED:11,TarekPhD,cmpt2012,igor:2012}, whereas only a limited number of publications have discussed nonlinear circuits~\cite{Tao:2007,Strunz:2008}. In~\cite{Tao:2007}, SC is combined with PC to simulate RF circuits with Gaussian variations. Later,~\cite{Strunz:2008} developed a SPICE-type stochastic simulator for nonlinear circuits. The key idea is to construct some stochastic library models for both linear and nonlinear devices by linearization and Galerkin projection. However, one has to reconstruct these library models for different uncertainty specifications and bias conditions, and thus industrial semiconductor device models cannot be feasibly integrated with this PC-based simulator.

There often exist non-Gaussian variations in practical circuit design, which cannot be easily handled by existing PC-based UQ tools. Due to the development of generalized Polynomial Chaos (gPC)~\cite{gPC2002,gPC2003,xiu2009}, spectral methods can now be applied to physical models with non-Gaussian variations, and extensive results have been reported~\cite{multibody:2006,Akil:2012,Akil:CSE,pulch_jcp,sgscCompare,book:Dxiu,xiu2009,sMOR2012,UQ:book}. Unfortunately, there seems to be limited research investigating the application of gPC to EDA problems. In~\cite{sMOR2012}, gPC was employed with SC to construct linear stochastic models for electromagnetic devices. Later, gPC-based SC and SG were applied to the UQ of linear circuits with Gaussian and non-Gaussian variations~\cite{pulch_jcp}. However, directly applying existing SG or SC methods to circuit problems can be inefficient, as will be discussed in Section IV and demonstrated by the examples in Section~\ref{sec:results}.

Among various SC methods, there exists a special kind of SC scheme~\cite{multibody:2006,Akil:2012,Akil:CSE,pulch_jcp}\footnote{The authors would like to thank the anonymous reviewer for pointing out the related work in the mathematical community, specifically Ref.~\cite{pulch_jcp,multibody:2006}.}. Different from the mainstream SC methods using sparse grids or tensor rules, this SC scheme uses the same number of basis functions and sampling nodes to construct a coupled deterministic equation. The resulting equation can be decoupled {\it a-priori} with a transformation~\cite{multibody:2006} and then solved by repeatedly calling existing deterministic solvers. Combined with gPC, this nonintrusive method has been used for the UQ of the nonlinear dynamic systems arising from multibody problems~\cite{multibody:2006} and of linear differential algebraic equations (DAEs) from linear circuit analysis~\cite{pulch_jcp}. In~\cite{pulch_jcp}, the tensor product rule is used to construct the basis functions and sampling nodes for SC, leading to some computational overhead.

\textbf{Our Contribution.} In this paper, we propose a gPC-based \textbf{intrusive} simulator, called \textbf{stochastic testing (ST)}, for the UQ of transistor-level simulation. This work is a variant of the interpolation-based SC~\cite{pulch_jcp,multibody:2006}. Our work uses a collocation testing method to set up a coupled equation, and decoupling is used to accelerate the computation. However, our ST simulator differs from the previous work in the following aspects:  
\begin{enumerate}

	\item Different from the nonintrusive SC in~\cite{pulch_jcp,multibody:2006}, our proposed method is an intrusive simulator: the resulting coupled equation is solved directly to obtain the spectral coefficients, \textbf{without} decoupling {\it a-priori}. To distinguish our simulator with the intrusive SG and nonintrusive sampling-based SC, we call it ``stochastic testing" (ST).
	
	\item ST uses fewer testing nodes than the mainstream SC algorithms~\cite{col:2005} (which use sampling nodes from tensor products or sparse grids) and the recent work in~\cite{pulch_jcp}, leading to remarkable computational speedup. ST provides extra speedup in time-domain simulation, since the intrusive nature of ST allows adaptive time stepping.
	
	\item Decoupling is applied \textbf{inside} the intrusive solver. This makes our solver much more efficient over existing intrusive solvers such as SG without sacrificing flexible time stepping controls. 
	
\end{enumerate}
Our algorithm is implemented in a SPICE-type stochastic simulator and integrated with several semiconductor device models for algorithm verification. The proposed method can be applied to many general engineering problems as the mathematical derivation is very generic and does not make any restrictive assumptions in the stochastic DAE's.

\textbf{Paper Organization.} In section II we review MC, the existing spectral methods for stochastic circuit simulation, as well as gPC. Section III presents our intrusive ST simulator and its numerical implementation. In Section IV, gPC-based SG and SC are briefly extended to nonlinear circuits and compared with ST, and we further classify various stochastic simulators. Section V provides some circuit simulation results and discusses the speedup of ST over SC in detail.

\section{Review: Stochastic Simulators and GPC}
Let us consider a general nonlinear circuit with random parameters. Applying modified nodal analysis (MNA)~\cite{mna:1975}, we obtain a stochastic Differential Algebraic Equation (DAE):
\begin{equation}
\label{eq:sdae}
\begin{array}{l}
 \displaystyle{\frac{{d\vec q\left( {\vec x\left( {t,\vec \xi } \right),\vec \xi } \right)}}{{dt}} }+ \vec f\left( {\vec x\left( {t,\vec \xi } \right),\vec \xi } \right) = B\vec u\left( t \right) 
 \end{array}
\end{equation}
where $\vec u(t)\in \mathbb{R}^m$ is the input signal, ${\vec x}\in \mathbb{R}^n$ denotes nodal voltages and branch currents, ${\vec q}\in \mathbb{R}^n$ and ${\vec f}\in \mathbb{R}^n$ represent the charge/flux term and current/voltage term, respectively. Vector ${\vec \xi}=[\xi_1; \xi_2; \;\cdots \;\xi_l]$ denotes $l$ random variables describing the device-level uncertainties assumed mutually independent. In this paper, the port selection matrix $B$ is assumed independent of the random parameters $\vec \xi$. We focus on how to solve~(\ref{eq:sdae}) to extract some statistical information such as mean, variance and probability density function ($\rm PDF$) of the state vector $\vec x\left( {t,\vec \xi } \right)$. 

\subsection{Monte Carlo Method}

Monte Carlo (MC) is the most widely used UQ tool, and it is implemented in almost all commercial circuit simulators. In MC, $N_s$ samples ${\vec \xi}^1, \cdots, {\vec \xi}^{N_s}$ are first generated according to ${\rm PDF}(\vec \xi)$, the joint Probability Density Function ($\rm PDF$) of ${\vec \xi}$. Any available deterministic solver is then called to run a simulation at each sample, generating a set of deterministic solutions. Finally, all deterministic solutions are utilized to compute the statistical characterization of interest. The error of MC is proportional to $\frac{1}{\sqrt{N_s}}$. Very often, a huge number (thousands to millions) of samples are required to obtain the desired level of accuracy even when improvements on sampling point selection, such as Mixture Importance Sampling, Quasi-Monte Carlo and Latin Hypercube sampling~\cite{Kanj:2006,SingheeR09,SingheeR10}, are used. The excessive number of samples render the repeated simulation prohibitively expensive in many cases.
\begin{table*}[t]
	\centering 
	\caption{Univariate gPC polynomial basis of different random parameters~\cite{xiu2009}.}	
	\label{tab:gPC}
	\begin{threeparttable}
	\begin{tabular}{|c||c|c|c|}
	\hline
Distribution of $\xi_k$ & ${\rm PDF}$ of component $\xi_k$ [$\rho_k(\xi_k)$]\footnotemark[1]& univariate gPC basis function $\phi _{i_k } \left( {\xi _k } \right)$ & Support of $\xi _k$\\
\thickhline	
	Gaussian  & $\frac{1}{{\sqrt {2\pi } }}\exp \left( {\frac{{ - \xi _k^2 }}{2}} \right)$ &Hermite-chaos polynomial&  $(-\infty, +\infty)$ \\ \hline
	Gamma  & $\frac{{\xi _k ^{\gamma  - 1} \exp \left( { - \xi _k } \right)}}{{\Gamma \left( \gamma  \right)}},\;\gamma  > 0$  &Laguerre-chaos polynomial&  $[0, +\infty)$ \\ \hline
	Beta  & $\frac{{ {\xi_k}^{\alpha  - 1} \left( {1 - \xi_k} \right)^{\beta  - 1} }}{{{\rm B}\left( {\alpha ,\beta } \right)}},\;\;\alpha ,\beta  > 0 $ &Jacobi-chaos polynomial&  $[0, 1]$ \\ \hline
	Uniform  & $\frac{1}{2}$  &Legendre-chaos polynomial&  $[-1, 1]$ \\	
\hline
	\end{tabular} 
	\begin{tablenotes}
       \item[1] $\Gamma \left( \gamma  \right) = \int\limits_0^\infty  {t^{\gamma  - 1} \exp \left( { - t} \right)dt}$ and ${\rm B}\left( {\alpha ,\beta } \right) = \int\limits_0^1 {t^{\alpha  - 1} \left( {1 - t} \right)^{\beta  - 1} dt}$ are the Gamma and Beta functions, respectively.\normalsize
  \end{tablenotes}
 \end{threeparttable}	 	
\end{table*}

\subsection{PC-based SG and SC Methods}
In the EDA community, most existing spectral stochastic simulators focus on linear circuits with Gaussian parameters~\cite{Tarek_DAC:08,Tarek_DAC:10,Tarek_ISQED:11,TarekPhD,cmpt2012,igor:2012} by considering the following linear stochastic DAE
\begin{equation}	
\label{ode}
E\left( {\vec \xi } \right)\frac{{d\vec x\left( t,{\vec \xi} \right)}}{{dt}} + A\left( {\vec \xi } \right)\vec x\left( t,{\vec \xi} \right) = Bu\left( t \right).
\end{equation}
Since ${\vec \xi}$ contains only Gaussian parameters, $\vec x\left( {t,\vec \xi } \right)$ can be well approximated by a truncated Hermite expansion
\begin{equation}
\label{eq:hermite}
\vec x\left( {t,\vec \xi } \right) \approx \sum\limits_{k = 1}^K {\hat x_{k } }(t) H_{k } ( {\vec \xi } )
\end{equation}
where $H_{k } ( {\vec \xi })$ is an orthonormal multivariate Hermite polynomial~\cite{sfem}, and ${\hat x_{k } }(t) $ the PC coefficient. If the total polynomial order is $p$, then the above Hermite expansion uses
\begin{equation}
\label{Kvalue}
K = \left( \begin{array}{l}
 p + l \\ 
 \;\;p 
 \end{array} \right)=\frac{(p+l)!}{p!l!}
\end{equation}
basis functions in total to approximate $\vec x({\vec \xi},t)$.

In the intrusive SG method~\cite{sfem}, the Hermite expansion~(\ref{eq:hermite}) is first substituted into (\ref{ode}). Applying Galerkin testing, SG sets up a coupled equation of dimension $nK$. The PC coefficients are then directly computed by solving this coupled equation. 

The nonintrusive SC method~\cite{col:2005} first selects a set of sampling points according to some rules (such as Gauss-quadrature tensor product rule or sparse grid rule). At each sampling point, (\ref{ode}) is solved as a deterministic equation to get a deterministic solution. After that, a post-processing step such as numerical integration is applied to get the PC coefficients. 

\subsection{Generalized Polynomial Chaos (gPC)}
Generalized polynomial chaos (gPC)~\cite{gPC2002,gPC2003,xiu2009} is a generalization of the original Hermite-type PC~\cite{PC1938}, and it can handle both Gaussian and non-Gaussian random parameters efficiently. A multivariate gPC basis function $H_{\vec i}(\vec \xi )$ reads
\begin{equation}
H_{\vec i} \left( {\vec \xi } \right) = \prod\limits_{k = 1}^l {\phi _{i_k } \left( {\xi _k } \right)}, 
\end{equation}
where $\phi _{i_k } \left( {\xi _k } \right)$ is a univariate orthonormal polynomial of degree $i_k $. The specific form of $\phi _{i_k } \left( {\xi _k } \right)$ depends on the density function of $\xi_k$. Table~\ref{tab:gPC} lists the correspondence between some typical univariate gPC polynomial basis $\phi _{i_k } \left( {\xi _k } \right)$ and the probability distributions of ${\xi _k }$~\cite{xiu2009}. 

In the stochastic space $\Omega$, the inner product of any two general functions $y_1 \left( {\vec \xi } \right)$ and $y_2 \left( {\vec \xi } \right)$ is defined as
\begin{equation}
\label{inner}
\left\langle {y_1 \left( {\vec \xi } \right),y_2 \left( {\vec \xi } \right)} \right\rangle =\int\limits_\Omega  {{\rm{PDF}}\left( {\vec \xi } \right)y_1 \left( {\vec \xi } \right)y_2 \left( {\vec \xi } \right)d\vec \xi }.
\end{equation}
The normalized gPC bases have the the orthogonality property
\begin{equation}
\begin{array}{l}
 \left\langle {H_{\vec i} \left( {\vec \xi } \right),H_{\vec j} \left( {\vec \xi } \right)} \right\rangle  = \delta_{\vec i,\vec j}. \nonumber
 \end{array}
\end{equation}
With gPC, one can also approximate a second-order stochastic process $\vec x({\vec \xi},t)$ by an order-$p$ truncated series
\begin{equation}	
\label{gpcExpan}
\vec x(t,\vec \xi ) \approx \sum\limits_{| {\vec i} | \le p} {\tilde x_{\vec i} (t)H_{\vec i}(\vec \xi )} 
\end{equation}
which has totally $K$ basis functions [$K$ is given in (\ref{Kvalue})]. In (\ref{gpcExpan}), 
${\vec i}$$=$$[i_1; i_2;\cdots; i_l]$ is the index vector with $| {\vec i} | $$= $$\sum\limits_{k = 1}^l {| {i_k } |}$, integer $i_k$ the highest order of $\xi_k$ in $H_{\vec i}(\vec \xi )$. 
The mean value and standard deviation of $\vec x(\vec \xi, t )$ are easily calculated as:
\begin{equation}	
\begin{array}{l}
 {\rm{E}}\left( {\vec x \left( t,{\vec \xi }\right)} \right) = \tilde x_{\vec i} (t),\;| {\vec i} | = 0 \\ 
 \sigma \left( {\vec x \left( t,{\vec \xi } \right)} \right) \approx \sqrt {\sum\limits_{| {\vec i} | = 1}^p {| {\tilde x_{\vec i} (t) } |^2 } }.  
 \end{array}
\end{equation}

In PC and gPC, the random parameters are assumed mutually independent. For general cases with arbitrary probability measures, constructing orthogonal basis functions is much more involved. A nice approach is proposed in~\cite{arb_chaos}, however, its numerical implementation is not trivial. In this paper, we keep the assumption that all random parameters are mutually independent, and we apply gPC to develop more efficient UQ tools for nonlinear transistor-level circuit analysis.

\section{Stochastic Testing Simulator}
Since there is a one-to-one correspondence between $1\leq k\leq K$ and the index vector $\vec i$, we denote
\begin{align}
\label{gpc:deno}
	\hat{x} ( {t,\vec \xi })=\sum\limits_{k = 1}^K {\hat x_{k } }(t) H_{k } ( {\vec \xi } )
\end{align}
for convenience. Now $H_{k } ( {\vec \xi } )$ denotes the $k$-th multivariate orthonormal gPC basis function of (\ref{gpcExpan}).
Replacing the exact solution $\vec x\left (t, \vec \xi \right)$ in stochastic DAE (\ref{eq:sdae}) with the above truncated gPC expansion yields a residual function
\begin{equation}
\label{eq:residual}
\begin{array}{l}
 {\rm Res}(\vec X(t),\vec \xi )=\displaystyle{\frac{{d\vec q\left( {\hat x( {t,\vec \xi } ),\vec \xi } \right)}}{{dt}}} + \vec f\left( {\hat x( {t,\vec \xi } ),\vec \xi } \right)- B\vec u( t ).
 \end{array}
\end{equation}
Now the unknown vector reads
\small
\begin{equation}
\vec X\left( t \right) =\left[\hat x_1 \left( t \right);\;\cdots ; \; \hat x_K \left( t \right)\right]  \in \mathbb{R}^N ,\;{\rm{with}}\;N = nK.
\end{equation}\normalsize

\subsection{Basic Idea of the ST Method}
In order to compute $\vec X \left( t \right)$, ST starts from (\ref{eq:residual}) and sets up a larger-size determined equation by collocation testing. Specifically, ST selects $K$ testing (or collocation) points $\vec \xi ^1, \cdots, \vec \xi ^K$, then it enforces the residual function to be zero at each point, leading to the following deterministic DAE:
\begin{align}
\label{label:dDAE}
\frac{{dQ\left( {\vec X\left( t \right)} \right)}}{{dt}} + F\left( {\vec X\left( t \right)} \right) = \tilde B \vec u\left( t \right)
\end{align}
with 
\begin{equation}
\label{eq:ST}
\begin{array}{l}
 Q\left( {\vec X\left( t \right)} \right) = \left[ {\begin{array}{*{20}c}
   {\vec q\left( \hat{x} ( {t,\vec \xi^1 }), \vec \xi^1   \right)}  \\
    \vdots   \\
   {\vec q\left( \hat{x} ( {t,\vec \xi^K }), \vec \xi^K   \right)}  \\
\end{array}} \right],\;\tilde B = \left[ {\begin{array}{*{20}c}
   {B}  \\
    \vdots   \\
   {B}  \\
\end{array}} \right] \\ 
 F\left( {\vec X\left( t \right)} \right) = \left[ {\begin{array}{*{20}c}
   {\vec f\left( \hat{x} ( {t,\vec \xi^1 }), \vec \xi^1   \right)}  \\
    \vdots   \\
   {\vec f\left( \hat{x} ( {t,\vec \xi^K }t), \vec \xi^K   \right)}  \\
\end{array}} \right]. \\ 
 \end{array}
\end{equation}

The collocation testing used in ST is the same with that used in collocation-based integral equation solvers~\cite{fastCap:1991}. However, in stochastic computation, ``stochastic collocation" means a different sampling-based method (c.f. Section~\ref{subsec:SC}). Therefore, we name our proposed method as ``stochastic testing".  

There remain two important issues, and how to address them distinguishes our ST solver with the nonintrusive stochastic solvers in~\cite{multibody:2006,pulch_jcp}. The first issue is how to solve the resulting coupled DAE. ST directly solves (\ref{label:dDAE}) by an intrusive solver. As a result, the gPC coefficients can be directly computed and adaptive time stepping~\cite{kundertbook:1995} can be used. The second issue is how to select the testing nodes. ST selects $K$ testing points from some candidate nodes, whereas $(p+1)^l$$\gg$$ K$ nodes are used in~\cite{pulch_jcp} to make the transformation matrix invertible.

\subsection{Intrusive Decoupled Solver}
\label{subsec:decoupled}
ST is an intrusive simulator: the coupled DAE is passed into a specialized transient solver to directly compute the gPC coefficients, and matrix structures are exploited inside Newton's iterations to obtain simulation speedup. As a demonstration, we consider backward-Euler integration. Other types of numerical integration schemes (e.g., Trapezoidal or Gear-2 method) are implemented in a similar way inside ST. 
\begin{figure}[t]
	\centering
		\includegraphics[width=50mm]{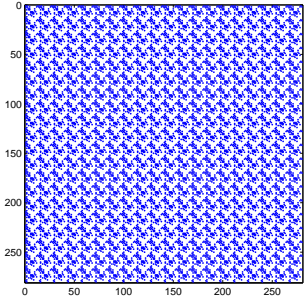} 
\caption{Structure of the Jacobian matrix in ST-based simulator, with $l=p=3$ and $K=20$.}
	\label{fig:stSparsity}
\end{figure}

Let $\vec X_k$$=$$\vec X\left( t_k \right)$ and $\vec u_k$$=$$\vec u\left( t_k \right)$. In the transient solver, DAE (\ref{label:dDAE}) is discretized, leading to an algebraic equation
\begin{align}
{\rm R}( {\vec X_k }) = \alpha_k ( {Q( {\vec X_k } ) - Q( {\vec X_{k - 1} } )} ) + F( {\vec X_k } ) - \tilde B\vec u_k  = 0 \nonumber
\end{align}
with $\alpha_k $$ = $$\frac{1}{{t_k  - t_{k - 1} }}$. The time step size is adaptively selected according to the local truncation error (LTE)~\cite{kundertbook:1995,Tuinenga:1995}. Starting from an initial guess $\vec X_k^0$, $\vec X_k $ is computed using Newton's iterations
\begin{align}
\label{newton}
\begin{array}{l}
 {\rm{solve}}\;{\cal J}\left( {\vec X_k^j } \right)\Delta \vec X_k^j  = -{\rm{R}}\left( {\vec X_k ^j} \right),\;\; \\ 
 {\rm{update}}\;\vec X_k^{j+1}  = \vec X_k^j  + \Delta \vec X_k^j, 
 \end{array}
\end{align}
until convergence. Here ${\cal J}( {\vec X_k^j } )$ is the Jacobian matrix of ${\rm R}( {\vec X_k^j } )$.  Fig.~\ref{fig:stSparsity} shows the structure of ${\cal J}( {\vec X_k^j } )$ from a CMOS low-noise amplifier (LNA) with $n$$=$$14$, $l$$=$$p$$=$$3$ and $K$$=$$20$. Clearly, all off-diagonal blocks are filled with non-zero submatrices. As a result, directly using a matrix solver to compute $\Delta \vec X_k^j $ can be inefficient. If a direct matrix solver is employed, the linear system solution costs $O(N^3)=O(K^3n^3)$; when an iterative method is applied, the cost is $\hat m O(K^2n)$ where $\hat m$ is the number of iterations.

The coupled linear equation in (\ref{newton}) is instead solved in a decoupled manner. We rewrite the Jacobian matrix in (\ref{newton}) as
\begin{align}
\label{Jacobian}
{\cal J}( {\vec X_k^j } ) = \tilde {\cal J}( {\vec X_k^j } ){ M}.
\end{align}
Matrix $\tilde {\cal J}( {\vec X_k^j } )$ has a block-diagonal structure: 
\begin{align}
\begin{array}{l}
 \tilde {\cal J}( {\vec X_k^j } ) = \left[ {\begin{array}{*{20}c}
   {J( \vec X_k^j ,\vec \xi ^1 )} & {} & {}  \\
   {} &  \ddots  & {}  \\
   {} & {} & {J( \vec X_k^j ,\vec \xi ^K )}  \\
\end{array}} \right].
 \end{array}
\end{align}
Let $\hat x_{n_2}^{k,j}$ denotes the $n_2$-th gPC coefficient vector in $X_k^j$, then
\begin{align}
\label{block:def}
J( \vec X_k^j ,\vec \xi ) = \left. { {\alpha_k \frac{{\partial \vec q( {\vec x,\vec \xi } )}}{{\partial \vec x}} + \frac{{\partial \vec f( {\vec x,\vec \xi } )}}{{\partial \vec x}}} } \right|_{\vec x = \sum\limits_{n_2 = 1}^K {\hat x_{n_2 }^{k,j} } H_{n_2 } ( {\vec \xi } )} .  
\end{align}
The matrix $M$ is
\begin{align}
\label{matrixKron}
M = \Phi  \otimes I_{n} ,\;\Phi  = \;\left[ {\begin{array}{*{20}c}
   {H_{1 } ( {\vec \xi ^1 } )} &  \cdots  & {H_{K } ( {\vec \xi ^1 } )}  \\
    \vdots  &  \ddots  & {\vdots}  \\
   {H_{1 } ( {\vec \xi ^K } )} & {} & {H_{K } ( {\vec \xi ^K } )}
\end{array}} \right]
\end{align}
where $\otimes$ denotes the Kronecker product operation. The Vandermonde-like matrix $\Phi\in \mathbb{R}^{K\times K}$ only depends on the testing points and basis functions. The inverse of $M$ is
\begin{align}
\label{Minverse}
M^{-1} = \Phi ^{ - 1}  \otimes I_{n \times n}
\end{align}
which can be easily computed because: 1) $\Phi$ is of small size; and 2) fast inverse algorithms exist for Vandermonde-like matrices~\cite{fastInverse}. Both $\Phi$ and $\Phi ^{ - 1} $ are calculated only once and then reused for all time points.

Finally, the linear equation in (\ref{newton}) is solved as follows:
\begin{enumerate}
	\item Solve $\tilde {\cal J}( {\vec X_k^j }  )\Delta z= -{\rm{R}}( {\vec X_k ^j} )$ for $\Delta z$. Due to the block-diagonal structure, this step costs only $KO\left(n^3\right)$ for a direct solver or $\hat m KO(n)$ for an iterative solver.
	\item Calculate the sparse matrix-vector product $\Delta \vec X_k^j=M^{-1}\Delta z$. Since the closed form of $M^{-1}$ is ready, the matrix-vector multiplication costs only $O(nK)$.
\end{enumerate}
The computational cost of ST solver now has only a linear dependence on $K$, as contrasted with the cubic or quadratic dependence when directly solving the coupled linear equation.

The ST solver can be easily implemented inside a commercial circuit simulator. Inside each Newton's iteration, one can convert $\vec X_k^j$ to a deterministic state variable and then evaluate the corresponding Jacobian and function values for a testing node. Repeating this procedure for all nodes, $\tilde {\cal J} (\vec X_k^j)$ and ${\rm R}(\vec X_k^j)$ can be obtained. After that, all blocks are solved independently to obtain $\Delta z$ and then $\Delta \vec X _k^j$. If the Newton's iterations get converged, the local truncation error (LTE) is checked by an existing estimator~\cite{kundertbook:1995,Tuinenga:1995}. The solution is accepted and ST proceeds to the next time point if the LTE is below a threshold; otherwise, the time step size is reduced and $\vec X_k$ is recomputed. Since the function/Jacobian evaluation and linear system solutions are well decoupled, ST can be easily implemented on a parallel computing platform.

\subsection{Testing Node Selection}
The testing nodes in ST are selected by two steps. First, $(p+1)^l$ candidate nodes are generated by a Gaussian-quadrature tensor product rule. Next, only $K$ nodes (with $K\ll p+1)^l$) are selected from the candidate nodes and used as the final testing nodes. Note that $(p+1)^l$ sampling nodes are used in~\cite{pulch_jcp}, which are exactly the candidate nodes of ST.

\subsubsection{Candidate Node Generation}
Let $\xi_k\in \Omega_k$ be a random parameter and $\rho_k (\xi_k)$ the corresponding PDF. Gaussian quadrature can be used to evaluate a 1-D stochastic integral:
\begin{equation}
\label{stoInt}
\int\limits_{\Omega _k } {g\left( {\xi _k } \right)\rho_k \left( {\xi _k } \right)d\xi _k }  \approx \sum\limits_{j = 1}^{\hat n} {g\left( {\xi _k^j } \right)} w_k^j
\end{equation}
where ${\xi _k^j }$ denotes the $j$-th quadrature point and $w_k^j$ the corresponding weight. The choice of a Gaussian quadrature rule depends on the support $\Omega_k$ and the PDF $\rho_k \left( {\xi _k } \right)$. 

With the computed 1-D quadrature points and weights for each $\xi_k$, one can construct multi-dimensional quadrature points to calculate the multivariate stochastic integral
\begin{equation}
\label{stoInt:md}
\int\limits_{\Omega } {g\left( {\vec \xi } \right){\rm PDF} \left( {\vec \xi } \right)d\vec \xi }  \approx \sum\limits_{j = 1}^{\hat N} {g\left( {\vec \xi _j } \right)} w^j
\end{equation}
by a tensor product or sparse grid technique~\cite{xiu2009,sgscCompare}. In this work, we set $\hat n=p+1$ [$p$ is highest total polynomial order in (\ref{gpcExpan})] and then use a tensor product rule to construct $\hat {N}=\hat n ^l$ quadrature nodes in the $l$-D stochastic space. For convenience, we define an index matrix ${\cal I}\in \mathbb{Z}^{l\times \hat N}$, the $j$-th column of which is decided according to the constraint 
\begin{equation}
\label{index}
1 + \sum\limits_{k = 1}^l {\left( {\hat n - 1} \right)^{k - 1} \left( {{\cal I}(k,j) - 1} \right)}=j. 
\end{equation}
Then the $j$-th  quadrature node in $\Omega$ is
\begin{align}
	\vec \xi _j  = [\xi _1^{{\cal I}\left( {1,j} \right)} , \cdots ,\xi _l^{{\cal I}\left( {l,j} \right)} ],
\end{align}
where $1\leq{\cal I}\left( {k,j} \right)\leq \hat n$ indicates the index of the quadrature point in $\Omega_k$. The corresponding weight of $\vec \xi _j$ is computed by
\begin{equation}
\label{weight}
{w} ^j  = \prod\limits_{k = 1}^l {w_k^{{\cal I}(k,j)}}. 
\end{equation} 

\subsubsection{Selecting Testing Nodes}
$K$ testing nodes are selected from the $(p+1)^l$ candidate nodes based on two criteria:
\begin{enumerate}
	\item We prefer those quadrature nodes that are statistically ``important", i.e., those nodes with large weight values;
	\item The matrix $\Phi$ should be full-rank and well conditioned.
\end{enumerate}
The Matlab pseudo codes of selecting the final testing nodes are provided in Algorithm~\ref{alg:testNode}. In Line $7$, $\beta>0$ is a threshold scalar.
The input vector in Line 2 is $ \vec w$$=$$[|w^1|,|w^2|, \cdots, |w^{\hat N}|]$, and the vector-valued function $\vec H(\xi)\in \mathbb{R}^{K\times 1}$ is
\begin{align}
	\vec H (\vec \xi)=[H_1(\vec \xi),H_2(\vec \xi), \cdots, H_K(\vec \xi)]^T.
\end{align}

The basic idea of Algorithm~\ref{alg:testNode} is as follows. All candidate nodes and their weights are reordered such that $|w^j|\geq |w^{j+1}|$, and the first node is selected as the first testing node $\vec \xi^1$. Then, we consider the remaining candidate nodes from the ``most important" to the ``least important". Assuming that $m-1$ testing nodes have been selected, this defines a vector space
\begin{align}
	V={\rm span} \left\{ {\vec H (\vec \xi^1),\cdots,\vec H (\vec \xi^{m-1})} \right\}.
\end{align}
The next ``most important" candidate $\vec \xi_k$ is selected as a new testing node if and only if $\vec H (\vec \xi_k)$ has a large enough component orthogonal to $V$. This means that the dimensionality of $V$ can be increased by adding $\vec \xi_k$ as a new testing point. 

\begin{algorithm}[t]
\caption{Testing Node Selection.}
\label{alg:testNode}
\begin{algorithmic}[1]
\STATE {Construct $\hat N$ $l$-D Gaussian quadrature nodes and weights; }
\STATE {[$\vec w, {\rm ind}$]=${\rm sort}$($\vec w$, `descend');   \;\; \% reorder the weights } 
\STATE {$V=\vec H\left( \vec \xi_k\right)/||\vec H\left( \vec \xi_k\right) ||$, with $k={\rm ind}(1)$; } 
\STATE {$\vec \xi^1=\vec \xi_k$, $m=1$;  \;\; \%      the 1st testing node} 
\STATE {\textbf{for} $j=2,\;\cdots$, $\hat N$ \textbf{do}}
 \STATE {\hspace{10pt} $k={\rm ind}(j)$, $\vec v=\vec H\left( \vec \xi_k\right)-V\left( V^T\vec H\left( \vec \xi_k\right)\right)$;}
 \STATE {\hspace{10pt}\textbf{if} $||\vec v||/||\vec H\left( \vec \xi_k\right) ||>\beta$}
   \STATE {\hspace{20pt}$V=[V;\vec v/||\vec v||]$, $m=m+1$ ;}
   \STATE {\hspace{20pt} $\vec \xi^m=\vec \xi_k$; \;\; \% select as a new testing node.}
    \STATE {\hspace{20pt}\textbf{if} $m\geq K$, break, \textbf{end};}
     \STATE {\hspace{10pt}\textbf{end if}}
     \STATE {\textbf{end for} } 
\end{algorithmic}
\end{algorithm}

When $l$ is large, generating and saving the candidate nodes and index matrix ${\cal I}$ become expensive. A solution is to select the testing nodes without explicitly generating the candidate nodes or ${\cal I}$. First, we generate weight $w^j$'s and the corresponding index $j$'s according to (\ref{weight}) and (\ref{index}), respectively. In the $k$-th step, we find the $k$-th largest weight $w^j$ and its corresponding index $j$. According to (\ref{index}), the $j$-th column of the index matrix ${\cal I}$ can be calculated, and then we can construct candidate node $\vec \xi_j$. Finally $\vec \xi_j$ is selected as a new testing node if $\vec H(\vec \xi_j)$ has a large enough component orthogonal to $V$, otherwise it is omitted and not stored.

There exist other possible ways to select the testing nodes. A recent progress is to generate the nodes by Leja sequences, a greedy approximation to Fekete nodes~\cite{Akil:CSE}. How to select the optimal testing nodes is still an open problem.

\section{Comparison with Other Stochastic Solvers}
This section briefly extends the gPC-based SG and SC to nonlinear circuit problems and compares them with our proposed ST algorithm. After that, a high-level classification of the mainstream stochastic solvers is presented.

\subsection{Comparison with Stochastic Galerkin (SG) Method}
\label{subsec:sg}
\subsubsection{SG for Nonlinear Circuits}
Similar to ST, SG starts from the residual function (\ref{eq:residual}), but it sets up a deterministic DAE in the form (\ref{label:dDAE}) by Galerkin testing. Specifically, SG enforces the residual function to be orthogonal to each gPC basis function:
\begin{equation}
\label{ode:galerkin}
\left\langle {{\rm Res}\left(\vec X(t),\vec \xi \right),H_{k } \left( {\vec \xi} \right)} \right\rangle  = 0, \; {\rm for}\; k=1,\cdots, K.
\end{equation}
Now $Q( \vec X( t))$, $F( \vec X( t))$ and $\tilde B$ in (\ref{label:dDAE}) have the block form
\begin{equation}
\label{eq:SG}
\begin{array}{l}
 Q\left( {\vec X( t )} \right) = \left[ {\begin{array}{*{20}c}
   {Q_1 \left( {\vec X( t )} \right)}  \\
    \vdots   \\
   {Q_K \left( {\vec X( t )} \right)}  \\
\end{array}} \right],\;\;\tilde B = \left[ {\begin{array}{*{20}c}
   {B_1 }  \\
    \vdots   \\
   {B_K }  \\
\end{array}} \right] \\ 
 F\left( {\vec X( t )} \right) = \left[ {\begin{array}{*{20}c}
   {F_1 \left( {\vec X( t )} \right)}  \\
    \vdots   \\
   {F_K \left( {\vec X( t )} \right)}  
\end{array}} \right] , 
 \end{array}
\end{equation}
with the $n_1$-th block defined by
\begin{align}	
\label{sg:inner}
\begin{array}{l}
 Q_{n_1} \left( {\vec X\left( t \right)} \right) = \left\langle {\vec q\left( {\hat x( {t,\vec \xi } ),\vec \xi } \right),H_{n_1} ( {\vec \xi } )} \right\rangle,  \\ 
 F_{n_1} \left( {\vec X\left( t \right)} \right) = \left\langle {\vec f\left( {\hat x( {t,\vec \xi } ),\vec \xi } \right),H_{n_1} ( {\vec \xi } )} \right\rangle,  \\ 
 B_{n_1}  = \left\langle {B,H_{n_1} ( {\vec \xi } )} \right\rangle.   
 \end{array}
\end{align}
To obtain the above inner product, one can use numerical quadrature or Monte Carlo integration~\cite{MCintro}.

\subsubsection{ST versus SG} 
Both of them are intrusive solvers, and the coupled DAEs from ST and SG have the same dimension. However, SG is much more expensive compared to ST.

First, SG must evaluate multivariate stochastic integrals, hence functions $\vec q$ and $\vec f$ must be evaluated at many quadrature or sampling nodes. This step is not cheap because evaluating a semiconductor device model (e.g., BISM3 model) at each node involves running tens of thousands of lines of codes.

Second, the linear system solution inside the Newton's iteration of SG is much more expensive. Assume that Gaussian quadrature is applied to calculate the inner products in (\ref{sg:inner}), then the Jacobian ${\cal J}( {\vec X }_k^j )$ has the following structure
\begin{align}
{\cal J}\left( {\vec X_k^j } \right) = \left[ {\begin{array}{*{20}c}
   {{\cal J}_{1,1} \left( {\vec X_k^j } \right)} &  \cdots  & {{\cal J}_{1,K} \left( {\vec X_k^j } \right)}  \\
    \vdots  &  \ddots  &  \vdots   \\
   {{\cal J}_{K,1} \left( {\vec X_k^j } \right)} &  \cdots  & {{\cal J}_{K,K} \left( {\vec X_k^j } \right)} 
\end{array}} \right],
\end{align}
and the submatrix ${\cal J}_{n_1,n_2}\left( {\vec X_k^j } \right) \in \mathbb{R}^{n\times n}$ is calculated by 
\begin{equation} 
\begin{array}{l}
 {\cal J}_{n_1 ,n_2 } \left( {\vec X_k^j } \right) 
 = \sum\limits_{q = 1}^{\hat N} {w^q H_{n_1 } \left( {\vec \xi ^q } \right)H_{n_2 } \left( {\vec \xi ^q } \right)} J\left( {\vec X_k^j ,\vec \xi ^q } \right).\nonumber
 \end{array}
\end{equation}
Here $\vec \xi^q$ is the $q$-th Gaussian quadrature node and $w^q$ the corresponding weight,  $J\left( {\vec X_k^j ,\vec \xi ^q } \right)$ is calculated according to the definition in (\ref{block:def}). The Jacobian in SG cannot be decoupled. Therefore, solving the resulting DAE of SG requires $O(N^3)=O(K^3n^3)$ at each time point if a direct solver is used (or $\hat m O(K^2n)$ if $\hat m$ iterations are used in an iterative solver), much more expensive compared to ST.

\begin{figure*}[t]
	\centering
		\includegraphics[width=160mm]{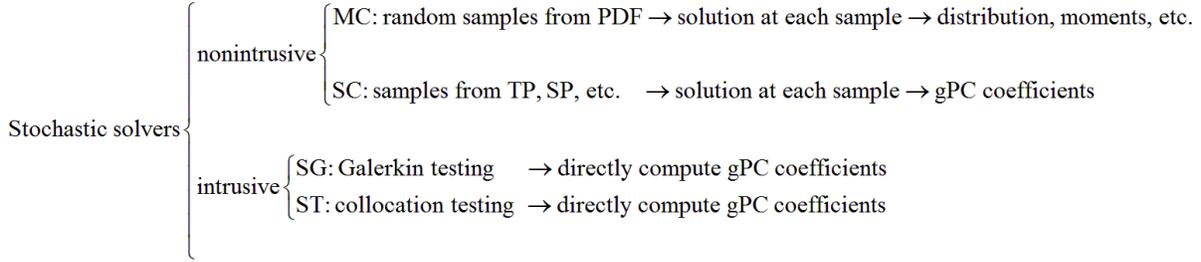} 
\caption{The classification of MC, SG, SC and ST methods.}
	\label{fig:method_class}
\end{figure*}  

\subsection{Comparison with Stochastic Collocation (SC) Method}
\label{subsec:SC}
\subsubsection{SC for Nonlinear Circuits}
Unlike ST and SG, SC starts from the original stochastic DAE (\ref{eq:sdae}) without using gPC approximation \textit{a-priori}. SC first selects $\hat N_s$ samples $\vec \xi ^1, \cdots, \vec \xi ^{\hat N_s}$ and solves (\ref{eq:sdae}) at each sample to obtain a deterministic solution $\vec x (t,\vec \xi^k)$. The gPC coefficients are then computed using a post-processing step. For example, one can select the sample $\vec \xi ^k$ and weight $w^k$ by a Gauss-quadrature tensor product rule or sparse grid technique, and then compute the gPC coefficient by
\begin{equation}
\label{SC:interpolation}
\hat x_{j } ( t ) = \left\langle {\vec x ( t,{\vec \xi } ),H_{j } ( {\vec \xi } )} \right\rangle\approx \sum\limits_{k = 1}^{\hat N_s } {w^k H_{j } ( {\vec \xi ^k } )} \vec x(t, {\vec \xi ^k } ).
\end{equation}

\subsubsection{ST versus SC} 
Like MC, SC is a sampling-based simulator. Therefore, the cost of SC has a linear dependence on $\hat N_s$, the number of samples. However, SC uses more sampling nodes than ST (c.f. Section~\ref{sec:expComp}). Furthermore, SC is not as efficient as ST in time-domain simulation. To reconstruct the gPC coefficients of time-domain solutions, SC must use the same time grid for simulating all deterministic DAEs. Since it is difficult to preselect an adaptive time grid, a small fixed step size is normally used, leading to excessive computational cost. In contrast, ST can use any standard adaptive step stepping to accelerate the time-domain simulation since it directly computes the gPC coefficients. It seems that SC can use adaptive time stepping to simulate each deterministic DAE, and then uses interpolation at the time points where solutions are missing. Unfortunately, the errors caused by such interpolations are much larger than the threshold inside Newton's iterations, causing inaccurate computation of higher-order gPC coefficients. However, SC indeed can use adaptive time stepping if one is not interested in the statistical information of the time-domain waveforms.

\subsection{Classification and Summary}
\begin{table}[t]
	\centering 
	\caption{Comparison of different spectral methods.}	
	\label{tab:compare}
	\begin{tabular}{|c||c|c|c|}
	\hline	
 Method &Type & Decoupled? & Adapt. step size?\\
\thickhline	
SC  &nonintrusive& $\surd$ & $\times$ \\ 
SG  &intrusive& $\times$ & $\surd$\\ 
ST  &intrusive&  $\surd$ & $\surd$ \\	
\hline
	\end{tabular} 	
\end{table}
\begin{figure}[t]
	\centering
		\includegraphics[width=45mm]{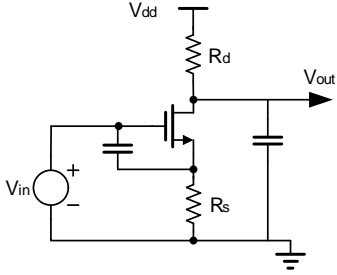} 
\caption{Schematic of the common-source amplifier.}
	\label{fig:Amp}
\end{figure}  

Fig.~\ref{fig:method_class} shows the classification of different stochastic solvers, which is detailed below. 

\begin{itemize}
	\item MC and SC are nonintrusive (or sampling-based) solvers. They both start from the original stochastic equation (\ref{eq:sdae}) and compute the deterministic solutions at a set of sampling points. The main difference of MC and SC lies in how to select the samples. MC draws the samples randomly according to ${\rm PDF}(\vec \xi)$, whereas SC selects the samples by a tensor-product (TP) numerical quadrature or sparse grid (SP) technique. After repeatedly simulating each deterministic equation, MC provides the statistical information such as distribution or moments, whereas SC reconstructs the gPC coefficients by a post-processing step such as numerical integration.
	
	\item SG and ST are intrusive solvers as they both directly compute the gPC coefficients by simulating a larger-size coupled DAE only once. With gPC approximations, they both start from the residual function (\ref{eq:residual}). SG sets up a larger-size coupled deterministic model by Galerkin testing, whereas ST uses a collocation testing technique. 
\end{itemize}

The spectral methods ST, SC and SG are further compared in Table~\ref{tab:compare}. ST allows both adaptive time stepping and decoupled simulation, therefore, it is more efficient over SC and SG.
\begin{figure}[t]
	\centering
		\includegraphics[width=80mm]{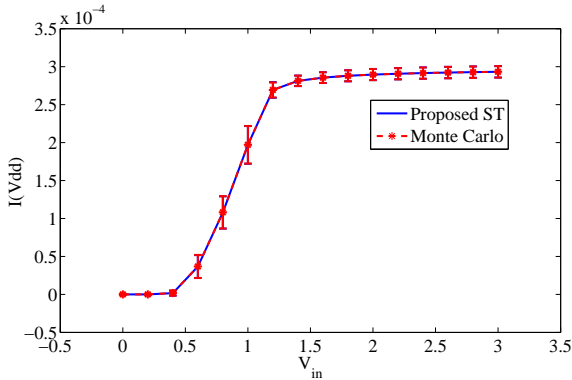} 
\caption{Error bars showing the mean and s.t.d values from our ST method (blue) and Monte Carlo method (red) of ${\rm I}({\rm V}_{{\rm dd}})$.}
	\label{fig:barAmp}
\end{figure}  
\begin{figure}[t]
	\centering
		\includegraphics[width=90mm]{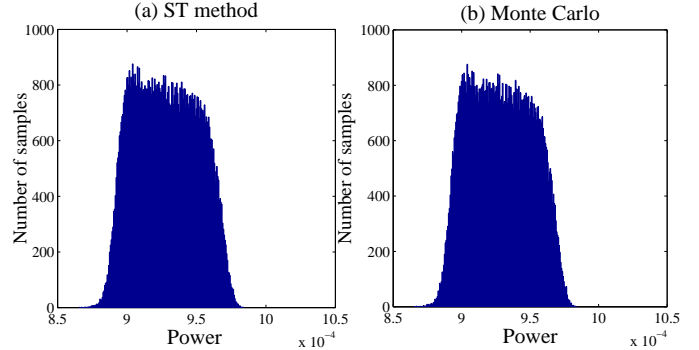}
\caption{Histograms showing the distributions of the power dissipation at $V_{\rm in}=1.4$V, obtained by ST method (left) and Monte Carlo (right).}
	\label{fig:histAmp}
\end{figure}

\section{Numerical Results}
\label{sec:results}
This section presents the simulation results of some analog, RF and digital integrated circuits. Our ST algorithm is implemented in a MATLAB prototype simulator and integrated with several semiconductor device models for algorithm verification. In this work, Level-3 MOSFET model and Ebers-Moll BJT model are used for transistor evaluation~\cite{starHSPICE}. The TSMC 0.25$\mu$m CMOS model card~\cite{mos25} is used to describe the device parameters of all MOSFETs. SC, SG and Monte Carlo (MC) methods are implemented for comparison and validation. In SG and ST, step sizes are selected adaptively according to the local truncation error (LTE)~\cite{kundertbook:1995} for time-domain simulation. In contrast, uniform step sizes are used for both MC and SC since we need to obtain the statistical information of time-domain solutions. In our experiments, all candidate nodes of ST are generated by Gaussian quadrature and tensor-product rules. The cost of generating the candidate nodes and selecting testing nodes is several milliseconds, which is negligible. For all circuit examples, SC and SG use the samples from a tensor-product rule. The sparse-grid and tensor-product SC methods are compared with ST in detail in Section~\ref{sec:expComp}.
\begin{figure}[t]
	\centering
		\includegraphics[width=90mm]{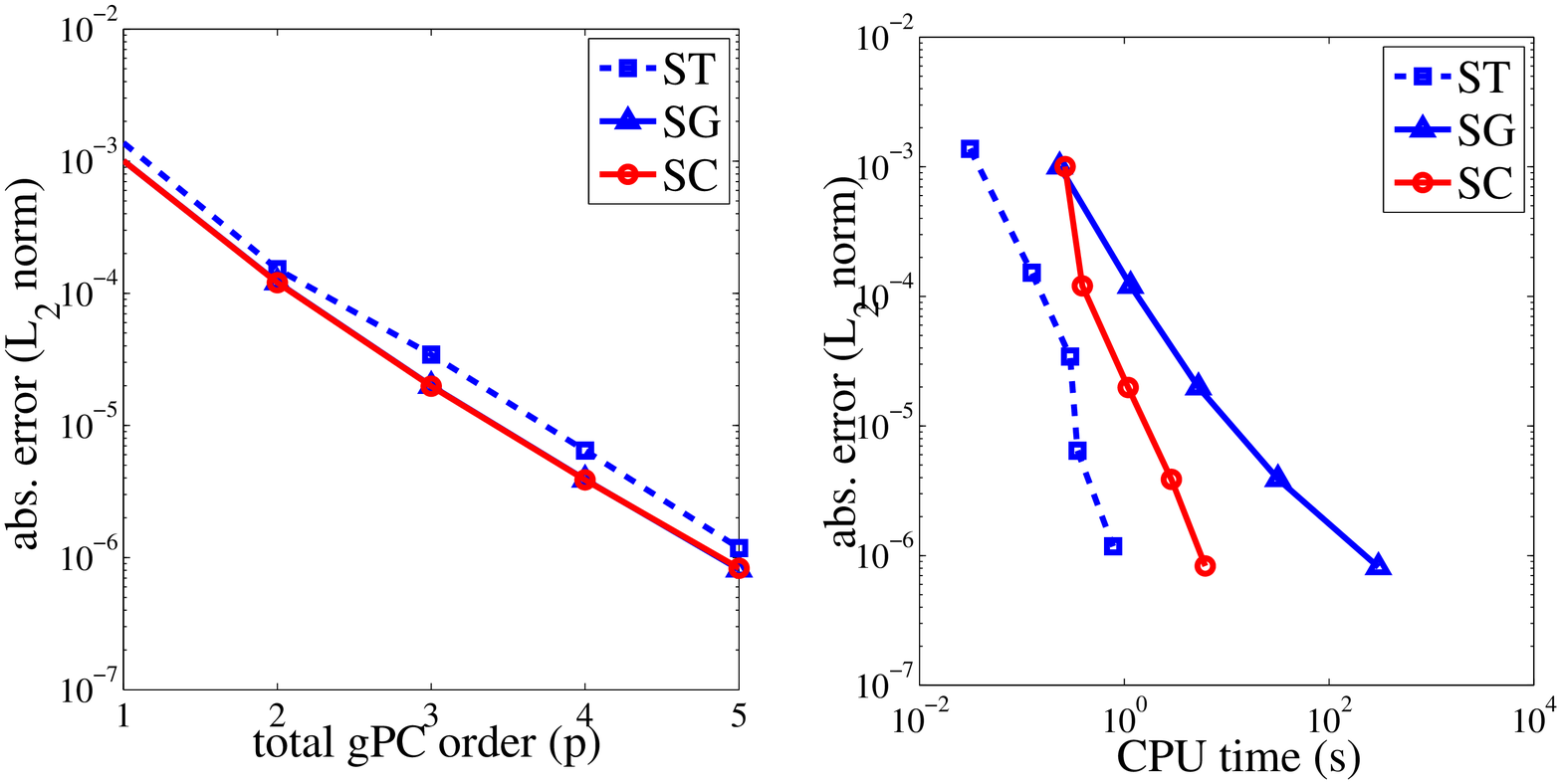} 
\caption{Absolute errors (measured by $L_2$ norm) of the gPC coefficients for the DC analysis of the CS amplifier, with $V_{\rm in}=1.6$V. Left: absolute errors versus gPC order $p$. Right: absolute errors versus CPU times.}
	\label{fig:methodCompare}
\end{figure}
\begin{table}[t]
	\centering 
	\caption{Computational cost of the DC analysis for CS amplifier.}	
	\label{tab:Timecompare}
	\begin{threeparttable}
	\begin{tabular}{|c|c||c|c|c|c|c|c|}
	\hline		
 \multicolumn{2}{|c||}{gPC order ($p$)} &$1$ & $2$ & $3$& $4$ & $5$ &$6$\\
\thickhline	
\multirow{2}{*}{ST} &time (s) &0.16& 0.22 & 0.29 & 0.51& 0.78 & 1.37 \\ \cline{2-8} 
 &$\# $ nodes &5& 15 & 35 & 70& 126 & 210 \\ \hline
\multirow{2}{*}{SC} & time (s)  &0.23& 0.33 & 1.09 & 2.89& 6.18 & 11.742 \\ \cline{2-8} 
 &$\# $ nodes &16& 81 & 256 & 625& 1296 & 2401 \\ \hline
\multirow{2}{*}{SG} & time (s)&0.25& 0.38 & 5.33 & 31.7& 304 & 1283 \\ \cline{2-8} 
&$\# $ nodes &16& 81 & 256 & 625& 1296 & 2401 \\ \hline
	\end{tabular} 
		\end{threeparttable}	
\end{table}

\subsection{Illustrative Example: Common-Source (CS) Amplifier}
\label{sec:csAmp}
The common-source (CS) amplifier in Fig.~\ref{fig:Amp} is used to compare comprehensively our ST-based simulator with MC and other spectral methods. This amplifier has $4$ random parameters: 1) ${\rm V}_{{\rm T}}$ (threshold voltage when ${\rm V}_{\rm bs}=0$) has a normal distribution; 2) temperate ${\rm T}$ has a shifted and scaled Beta distribution, which influences ${\rm V}_{\rm th}$; 3) ${\rm R}_{\rm s}$ and ${\rm R}_{\rm d}$ have Gamma and uniform distributions, respectively.
\begin{figure}[t]
	\centering
		\includegraphics[width=85mm]{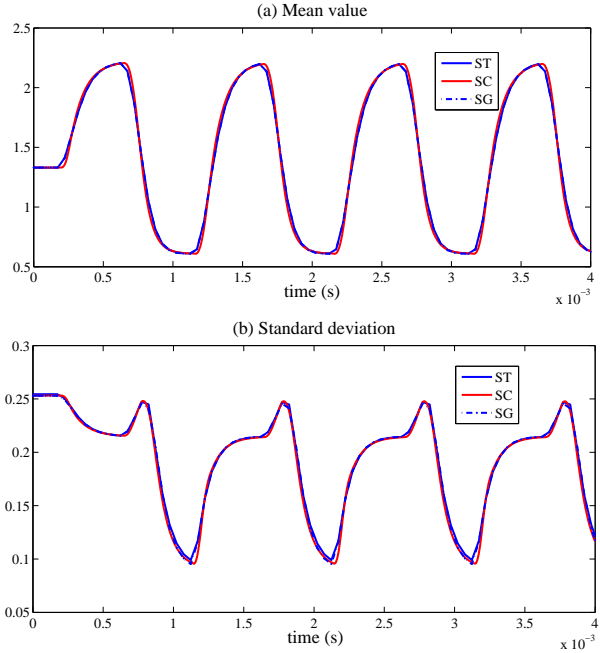} 
\caption{Transient waveform of the output of the CS amplifier.}
	\label{fig:amp_Wave}
\end{figure}
\begin{table}[t]
	\centering 
	\caption{Computational cost of transient simulation for CS amplifier.}	
	\label{tab:time_trans_cs}
	\begin{tabular}{|c||c|c|c|}
	\hline
 Methods &ST &SG  &SC\\
\thickhline	
 CPU times &41 s & $>1$ h & $1180$ s \\	\hline
 $\#$ nodes &35 & 256 & 256 \\	\hline
 speedup of ST & 1 & $>88$ & 29\\
\hline
	\end{tabular} 	
\end{table}  
\subsubsection{ST versus MC} ST method is first compared with MC in DC sweep. By sweeping the input voltage from $0$ V up to $3$ V with a step size of $0.2$ V, we estimate the supply currents and DC power dissipation. In MC, $10^5$ sampling points are used. In our ST simulator, using an order-$3$ truncated gPC expansion (with 35 gPC basis functions, and 35 testing nodes selected from 256 candidate nodes) achieves the same level of accuracy. The error bars in Fig.~\ref{fig:barAmp} show that the mean and ${\rm s.t.d}$ values from both methods are indistinguishable.
The histograms in Fig.~\ref{fig:histAmp} plots the distributions of the power dissipation at $V_{\rm in}=1.4$V. Again, the results obtained by ST is consistent with MC. The expected value at $1.4$V is $0.928$ mW from both methods, and the s.t.d. value is $22.07$ $\mu$W from both approaches. Apparently, the variation of power dissipation is not a Gaussian distribution due to the presence of circuit nonlinearity and non-Gaussian random parameters.

\textbf{CPU times:} For this DC sweep, MC costs about $2.6$ hours, whereas our ST simulator only costs $5.4$ seconds. Therefore, a $1700\times$ speedup is achieved by using our ST simulator.

\subsubsection{ST versus SC and SG in DC Analysis} Next, ST method is compared with SG and SC. Specifically, we set $V_{\rm in}=1.6$V and compute the gPC coefficients of all state variables with the total gPC order $p$ increasing from $1$ to $6$. We use the results from $p=6$ as the ``exact solution" and plot the $L_2$ norm of the absolute errors of the computed gPC coefficients versus $p$ and CPU times, respectively. The left part of Fig.~\ref{fig:methodCompare} shows that as $p$ increases, ST, SC and SG all converge very fast. Although ST has a slightly lower convergence rate, its error still rapidly reduces to below $10^{-4}$ when $p= 3$. The right part of Fig.~\ref{fig:methodCompare} shows that ST costs the least CPU time to achieve the same level of accuracy with SC and SG, due to the decoupled Newton's iterations and fewer nodes used in ST. 

\textbf{CPU times:} The computational costs of different solvers are summarized in Table~\ref{tab:Timecompare}. The speedup of ST becomes more significant as the total gPC order $p$ increases. We remark that the speedup factor will be smaller if SC uses sparse grids, as will be discussed in Section~\ref{sec:expComp}.  
\begin{figure}[t]
	\centering
		\includegraphics[width=70mm]{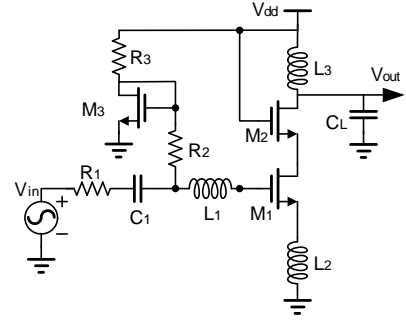}
\caption{Schematic of the LNA.}
	\label{fig:LNA}	
\end{figure}  
\begin{figure}[t]
	\centering
		\includegraphics[width=90mm]{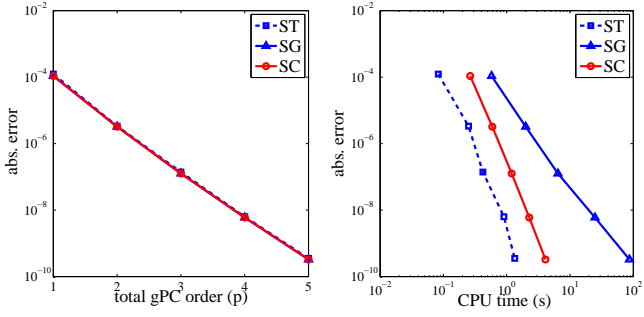} 
\caption{Absolute errors (measured by $L_2$ norm) of the gPC coefficients for the DC analysis of LNA. Left: absolute errors versus gPC order $p$. Right: absolute errors versus CPU times.}
	\label{fig:methodCompare_LNA}
\end{figure}

\subsubsection{ST versus SC and SG in Transient Simulation} Finally, ST is compared with SG and SC in transient simulation. It is well known that the SG method provides an optimal solution in terms of accuracy~\cite{gPC2002,gPC2003,xiu2009}, therefore, the solution from SG is used as the reference for accuracy comparison. The total gPC order is set as $p=3$ (with $K=35$ testing nodes selected from 256 candidate nodes), and the Gear-2 integration scheme~\cite{kundertbook:1995} is used for all spectral methods. In SC, a uniform step size of $10 \mu$s is used, which is the largest step size that does not cause simulation failures. The input is kept as $V_{\rm in}=1$ V for $0.2$ ms and then added with a small-signal square wave (with $0.2$V amplitude and $1$ kHz frequency) as the AC component. The transient waveforms of $V_{\rm out}$ are plotted in Fig.~\ref{fig:amp_Wave}. The mean value and standard deviation from ST are almost indistinguishable with those from SG. 

It is interesting that the result from ST is more accurate than that from SC in this transient simulation example. This is because of the employment of LTE-based step size control~\cite{kundertbook:1995}. With a LTE-based time stepping~\cite{kundertbook:1995}, the truncation errors caused by numerical integration can be well controlled in ST and SG. In contrast, SC cannot adaptively select the time step sizes according to LTEs, leading to larger integration errors.

\textbf{CPU times:} The computational costs of different solvers are summarized in Table~\ref{tab:time_trans_cs}. It is noted that SC uses about $7\times$ of nodes of ST, but the speedup factor of ST is $29$. This is because the adaptive time stepping in ST causes an extra speedup factor of about $4$. MC is prohibitively expensive for transient simulation and thus not compared here.

\subsection{Low-Noise Amplifier (LNA)}
\begin{table}[t]
	\centering 
	\caption{Computational cost of the DC analysis for LNA.}	
	\label{tab:Timecompare_LNA}
	\begin{threeparttable}
	\begin{tabular}{|c|c||c|c|c|c|c|c|}
	\hline
 \multicolumn{2}{|c||}{gPC order ($p$)} &$1$ & $2$ & $3$& $4$ & $5$ &$6$\\
\thickhline	
\multirow{2}{*}{ST} &time (s) &0.24& 0.33 & 0.42 & 0.90& 1.34 & 2.01 \\ \cline{2-8} 
 &$\# $ nodes &4& 10 & 20 & 35& 56 & 84 \\ \hline
\multirow{2}{*}{SC} & time (s)  &0.26& 0.59 & 1.20 & 2.28& 4.10 & 6.30 \\ \cline{2-8} 
 &$\# $ nodes &8& 27 & 64 & 125& 216 & 343 \\ \hline
\multirow{2}{*}{SG} & time (s) &0.58& 2.00 & 6.46 & 24.9& 87.2 & 286 \\ \cline{2-8} 
 &$\# $ nodes &8& 27 & 64 & 125& 216 & 343 \\ \hline
	\end{tabular} 
		\end{threeparttable}	
\end{table}
\begin{figure}[t]
	\centering
		\includegraphics[width=85mm]{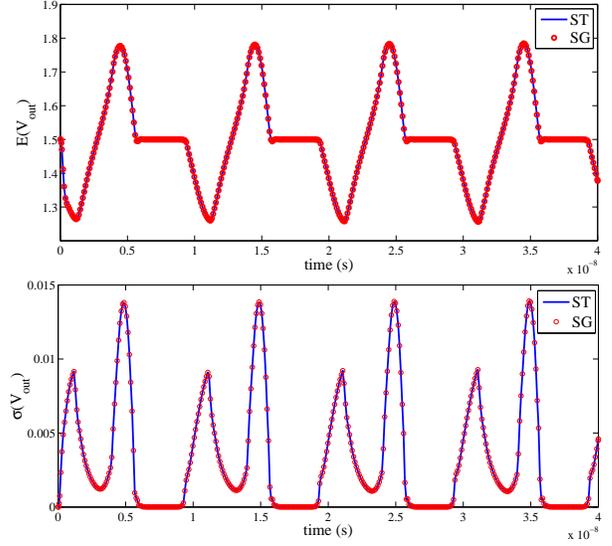} 
\caption{Transient simulation results of the LNA. Upper part: expectation of the output voltage; bottom part: standard deviation of the output voltage.}
	\label{fig:LNA_wave}
\end{figure}  

Now we consider a practical low-noise amplifier (LNA) shown in Fig~\ref{fig:LNA}. This LNA has $3$ random parameters in total: resistor $R_{3}$ is a Gamma-type variable; $R_{2}$ has a uniform distribution; the gate width of ${\rm M}_1$ has a uniform distribution. 

\textbf{DC Analysis:} We first run DC analysis by ST, SC and SG with $p$ increasing from $1$ to $6$, and plot the errors of the gPC coefficients of the state vector versus $p$ and CPU times in Fig.~\ref{fig:methodCompare_LNA}. For this LNA, ST has almost the same accuracy with SC and SG, and it requires the smallest amount of CPU time. The cost of the DC analysis is summarized in Table~\ref{tab:Timecompare_LNA}.

\textbf{Transient Analysis:} An input signal $V_{\rm in}=0.5{\rm sin}(2\pi ft)$ with $f=10^8$ Hz is added to this LNA. We are interested in the uncertainties of the transient waveform at the output. Setting $p=3$, our ST method uses $20$ gPC basis functions (with 20 testing nodes selected from 64 candidate nodes) to obtain the waveforms of the first $4$ cycles. The result from ST is indistinguishable with that from SG, as shown in Fig.~\ref{fig:LNA_wave}. ST consumes only $56$ seconds for this LNA. Meanwhile, SG costs $26$ minutes, which is $28\times$ slower compared to ST.

\subsection{6-T SRAM Cell}
\label{sec:sram}
\begin{figure}[t]
	\centering
		\includegraphics[width=65mm]{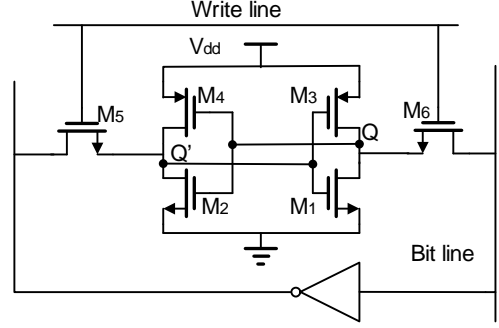} 
\caption{Schematic of the CMOS 6-T SRAM.}
	\label{fig:SRAM}
\end{figure} 
\begin{figure}[t]
	\centering
		\includegraphics[width=90mm]{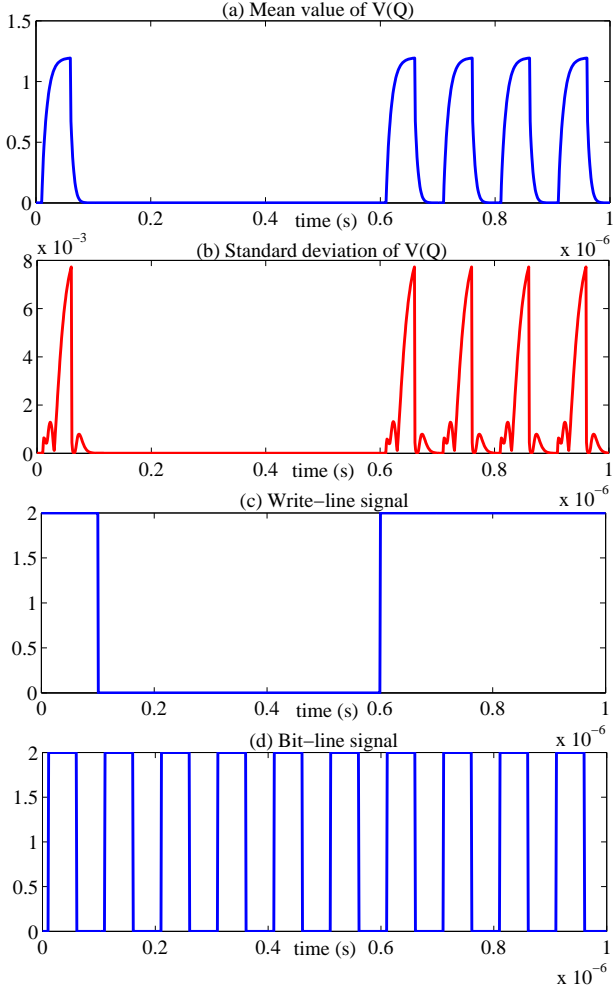} 
\caption{Uncertainties of the SRAM cell. (a) and (b) shows the expectation and standard deviation of $V_{\rm out}$; (c) and (d) shows the waveforms of the write line and bit line, respectively.}
	\label{fig:SRAM_results}
\end{figure}  

The 6-T SRAM cell in Fig.~\ref{fig:SRAM} is studied to show the application of ST in digital cell analysis. When the write line has a high voltage (logic 1), the information of the bit line can be written into the cell and stored on transistors ${\rm M}_1 - {\rm M}_4$. The 1-bit information is represented by the voltage of node ${\rm Q}$. When the write line has a low voltage (logic 0), ${\rm M}_5$ and ${\rm M}_6$ turn off. In this case, ${\rm M}_1 -{\rm M}_4$ are disconnected with the bit line, and they form a latch to store and hold the state of node ${\rm Q}$. Here $V_{\rm dd}$ is set as $1$ V, while the high voltages of the write and bit lines are both set as $2$ V.

Now we assume that due to mismatch, the gate widths of ${\rm M}_1 - {\rm M}_4$ have some variations which can be expressed as Gaussian variables. Here we study the influence of device variations on the transient waveforms, which can be further used for power and timing analysis. Note that in this paper we do not consider the rare failure events of SRAM cells~\cite{Kanj:2006}. To quantify the uncertainties of the voltage waveform at node ${\rm Q}$, our ST method with $p=3$ and $K=35$ (with 35 testing nodes selected from 256 candidate nodes) is applied to perform transient simulation under a given input waveforms. Fig.~\ref{fig:SRAM_results} shows the waveforms of write and bit lines and the corresponding uncertainties during the time interval $[0,1]\mu$s. 

\textbf{CPU times:} Our ST method costs $6$ minutes to obtain the result. SG generates the same results at the cost of several hours. Simulating this circuit with SC or MC is prohibitively expensive, as a very small uniform step size must be used due to the presence of sharp state transitions.
\begin{figure}[t]
	\centering
		\includegraphics[width=70mm]{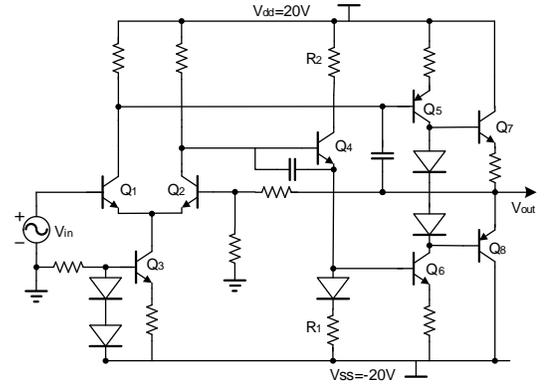}
\caption{Schematic of the BJT feedback amplifier.}
	\label{fig:bjtAmp}
\end{figure} 
\begin{figure}[t]
	\centering
		\includegraphics[width=80mm]{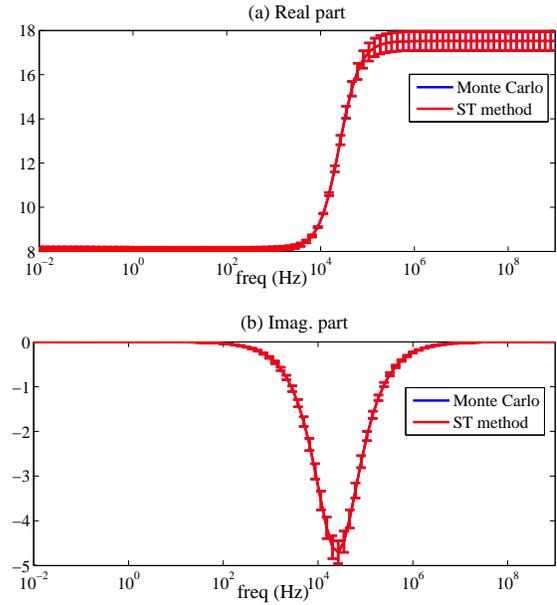}
\caption{Uncertainties of the transfer function of the BJT amplifier.}
	\label{fig:bjtAmp_tf}
\end{figure}  
\begin{figure}[t]
	\centering
		\includegraphics[width=80mm]{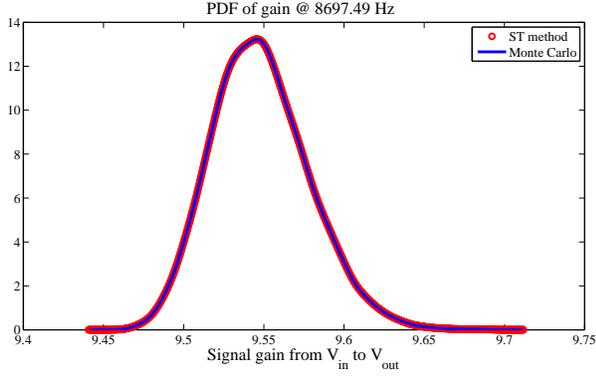} 
\caption{Simulated probability density functions of the signal gain.}
	\label{fig:bjtAmp_pdf}
\end{figure} 

\subsection{BJT Feedback Amplifier}
To show the application of our ST method in AC analysis and in BJT-type circuits, we consider the feedback amplifier in Fig.~\ref{fig:bjtAmp}. In this circuit, $R_1$ and $R_2$ have Gamma-type uncertainties. The temperature is a Gaussian variable which significantly influences the performances of BJTs and diodes. Therefore, the transfer function from $V_{\rm in}$ to $V_{\rm out}$ is uncertain. 

Using $p=3$ and $K=20$ (with 20 testing nodes selected from 64 candidate nodes), our ST simulator achieves the similar level of accuracy of a MC simulation using $10^5$ samples. The error bars in Fig.~\ref{fig:bjtAmp_tf} show that the results from both methods are indistinguishable. In ST, the real and imaginary parts of the transfer functions are both obtained as truncated gPC expansions. Therefore, the signal gain at each frequency point can be easily calculated with a simple polynomial evaluation. Fig.~\ref{fig:bjtAmp_pdf} shows the calculated PDF of the small-signal gain at $f=8697.49$ Hz using both ST and MC. The PDF curves from both methods are indistinguishable.

\textbf{CPU times:} The simulation time of ST and Monte Carlo are $3.6$ seconds and over $2000$ seconds, respectively.

\subsection{BJT Double-Balanced Mixer}
\label{sec:dbmixer}
\begin{figure}[t]
	\centering
		\includegraphics[width=75mm]{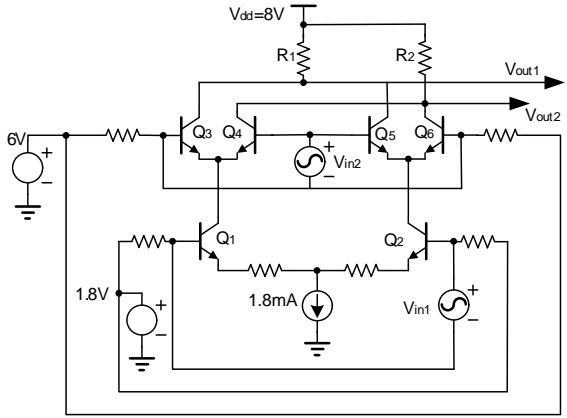} 
\caption{Schematic of the BJT double-balanced mixer.}
	\label{fig:dbmixer}
\end{figure}

\begin{figure}[t]
	\centering
		\includegraphics[width=85mm]{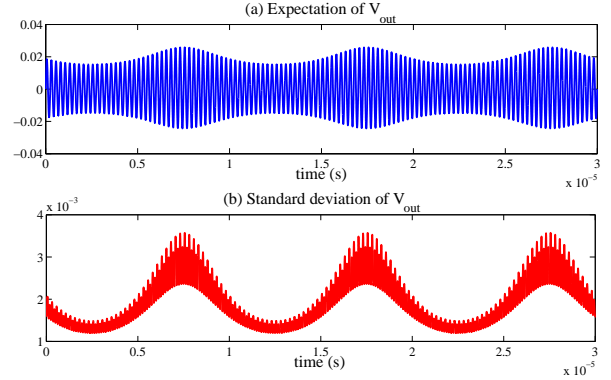} 
\caption{Uncertainties of $V_{\rm out}$$=$$V_{{\rm out}1}-V_{{\rm out}2}$ of the double-balanced mixer.}
	\label{fig:dbmixer_output}
\end{figure}

As the final circuit example, we consider the time-domain simulation of RF circuits excited by multi-rate signals, by studying the double-balanced mixer in Fig.~\ref{fig:dbmixer}. Transistors ${\rm Q}_1$ and ${\rm Q}_2$ accept an input voltage of frequency $f_1$, and ${\rm Q}_3\sim {\rm Q}_6$ accept the second input of frequency $f_2$. The output $v_{\rm out}=V_{\rm out1}-V_{\rm out2}$ will have components at two frequencies: one at $|f_1-f_2|$ and the other at $f_1+f_2$. Now we assume that $R_1$ and $R_2$ are both Gaussian-type random variables, and we measure the uncertainties of the output voltage. In our simulation, we set $V_{{\rm in}1}=0.01{\rm sin}(2\pi f_1 t)$ with $f_1=4$ MHz and $V_{{\rm in}2}=0.01{\rm sin}(2\pi f_2 t)$ with $f_2=100$ kHz. We set $p=3$ and $K=10$ (with 10 testing nodes selected from 16 candidate nodes), and then use our ST simulator to run a transient simulation from $t= 0$ to $t= 30\mu$s. The expectation and standard deviation of $V_{{\rm out}1}-V_{{\rm out}2}$ are plotted in Fig.~\ref{fig:dbmixer_output}. 

\textbf{CPU times:} The cost of our ST method is 21 minutes, whereas simulating this mixer by SG, SC or MC on the same MATLAB platform is prohibitively expensive due to the presence of multi-rate signals and the large problem size.

\subsection{Discussion: Speedup Factor of ST over SC}
\label{sec:expComp}
Finally we comprehensively compare the costs of ST and SC. Two kinds of SC methods are considered according to the sampling nodes used in the solvers~\cite{sgscCompare}: SC using tensor product (denoted as SC-TP) and SC using sparse grids (denoted as SC-SP). SC-TP uses $(p+1)^l$ nodes to reconstruct the gPC coefficients, and the work in~\cite{pulch_jcp} belongs to this class. For SC-SP, a level-$p+1$ sparse grid must be used to obtain the $p$-th-order gPC coefficients in (\ref{SC:interpolation}). We use the Fej\`{e}r nested sparse grid in~\cite{UQ:book}, and according to~\cite{sparsegrids} the total number of nodes in SC-SP is estimated as
\begin{align}
N_{{\rm{SC - SP}}}  = \sum\limits_{i = 0}^p {2^i \frac{{(l - 1 + i)!}}{{(l - 1)!i!}}} 
\end{align}
\begin{figure}[t]
	\centering
		\includegraphics[width=90mm]{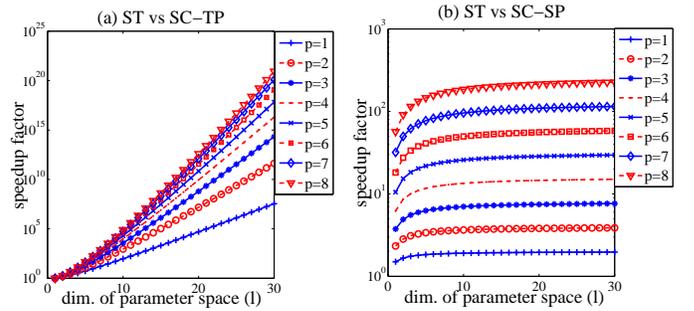} 
\caption{The speedup factor of ST over SC caused by node selection: (a) ST versus SC-TP, (b) ST versus SC-SP. This is also the speedup factor in DC analysis.}
	\label{fig:SPfactor}
\end{figure} 

\textbf{DC Analysis:} In DC analysis, since both ST and SC use decoupled solvers and their costs linearly depend on the number of nodes, the speedup factor of ST versus SC is 
\begin{align}
	\nu_{\rm DC}\approx N_{\rm SC} /K 
\end{align}
where $N_{\rm SC}$ and $K$ are the the numbers of nodes used by SC and ST, respectively. Fig.~\ref{fig:SPfactor} plots the values of $N_{\rm SC} /K$ for both SC-TP and SC-SP, which is also the speedup factor of ST over SC in DC analysis. Since ST uses the smallest number of nodes, it is more efficient over SC-TP and SC-SP. When low-order gPC expansions are used ($p\leq 3$), the speedup factor over SC-SP is below $10$. The speedup factor can be above $10$ if $p\geq 4$, and it gets larger as $p$ increases. In high-dimensional cases ($l\gg 1$), the speedup factor of ST over SC-SP only depends on $p$. It is the similar case if Smolyak sparse grids are used in SC~\cite{col:2005}. For example, compared with the sparse-grid SC in~\cite{col:2005}, our ST has a speedup factor of $2^p$ if $l\gg 1$.

\textbf{Transient Simulation:} The speedup factor of ST over SC in a transient simulation can be estimated as
\begin{align}
	\nu_{\rm Trans}\approx (N_{\rm SC} /K) \times \kappa, \; {\rm with} \; \kappa>1,
\end{align}
which is larger than $\nu_{\rm DC}$. The first part is the same as in DC analysis. The second part $\kappa$ represents the speedup caused by adaptive time stepping in our intrusive ST simulator, which is case dependent. For weakly nonlinear analog circuits (e.g., the SC amplifier in Section~\ref{sec:csAmp}), $\kappa$ can be below $10$. For digital cells (e.g., the SRAM cell in Section~\ref{sec:sram}) and multi-rate RF circuits (e.g., the double-balanced mixer in Section~\ref{sec:dbmixer}), SC-based transient simulation can be prohibitively expensive due to the inefficiency of using a small uniform time step size. In this case, $\kappa$ can be significantly large.

\section{Conclusion}
This paper has proposed an intrusive-type stochastic solver, named stochastic testing (ST), to quantify the uncertainties in transistor-level circuit analysis. With gPC expansions, ST can handle both Gaussian and non-Gaussian variations. Compared with SG and SC, ST can simultaneously allow decoupled numerical simulation and adaptive step size control. In addition, multivariate integral calculation is avoided in ST. Such properties make ST method hundreds to thousands of times faster over Monte Carlo, and tens to hundreds of times faster than SG. The speedup of ST over SC is caused by two factors: 1) a smaller number of nodes required in ST; and 2) adaptive time stepping in the intrusive ST simulator. The overall speedup factor of ST over SC is normally case dependent. Various simulations (e.g., DC, AC and transient analysis) are performed on some analog, digital and RF circuits, demonstrating the effectiveness of our proposed algorithm.

%

\bibliographystyle{IEEEtran}
\bibliography{ST} 
\begin{IEEEbiography}[{\includegraphics[width=1in,height=1.25in,clip,keepaspectratio]{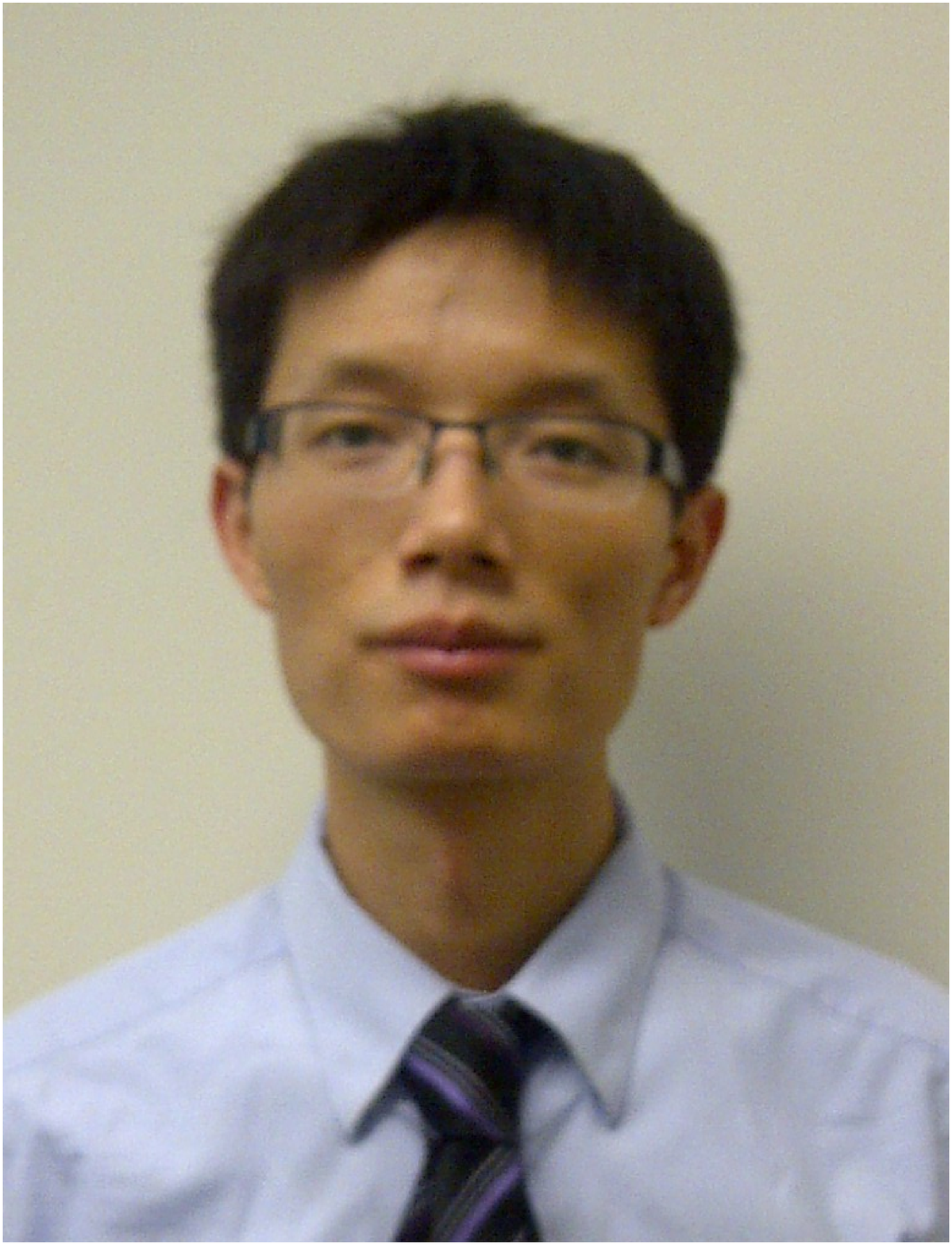}}]{Zheng Zhang} (S'09) received his B.Eng. degree from Huazhong University of Science and Technology, China, in 2008, and M.Phil. degree from the University of Hong Kong, Hong Kong, in 2010. He is a Ph.D student in Electrical Engineering at the Massachusetts Institute of Technology (MIT), Cambridge, MA. His research interests include uncertainty quantification, numerical methods for the computer-aided design (CAD) of integrated circuits and microelectromechanical systems (MEMS), and model order reduction.

In 2009, Mr. Zhang was a visiting scholar with the University of California, San Diego (UCSD), La Jolla, CA. In 2011, he collaborated with Coventor Inc., working on CAD tools for MEMS design. He was recipient of the Li Ka Shing Prize (university best M.Phil/Ph.D thesis award) from the University of Hong Kong, in 2011, and the Mathworks Fellowship from MIT, in 2010.
\end{IEEEbiography}

\begin{IEEEbiography}[{\includegraphics[width=1in,height=1.25in,clip,keepaspectratio]{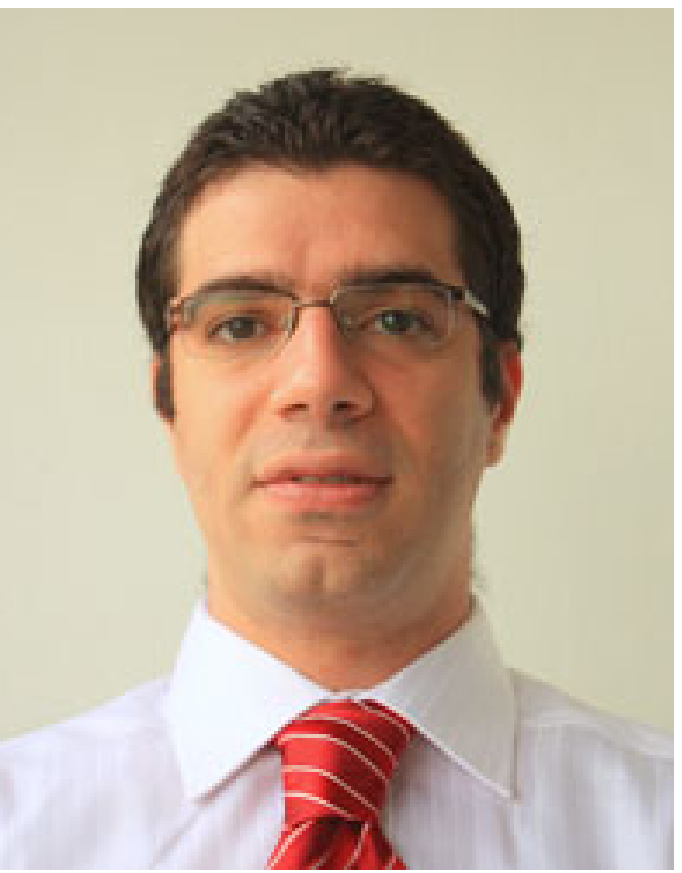}}]{Tarek El-Moselhy} received the B.Sc. degree in electrical engineering in 2000 and a diploma in mathematics in 2002, then the M.Sc. degree in mathematical engineering, in 2005, all from Cairo University, Cairo, Egypt. He received the Ph.D.
degree in electrical engineering from Massachusetts Institute of Technology, Cambridge, in 2010.

He is a postdoctoral associate in the Department of Aeronautics and Astronautics at Massachusetts Institute of Technology (MIT). His research interests include fast algorithms for deterministic and stochastic electromagnetic simulations, stochastic algorithms for uncertainty quantification in high dimensional systems, and stochastic inverse problems with emphasis on Bayesian inference. Dr. El-Moselhy received the Jin Au Kong Award for Outstanding PhD Thesis in Electrical Engineering from MIT in 2011, and the IBM Ph.D Fellowship in 2008.
\end{IEEEbiography}

\begin{IEEEbiography}[{\includegraphics[width=1in,height=1.25in,clip,keepaspectratio]{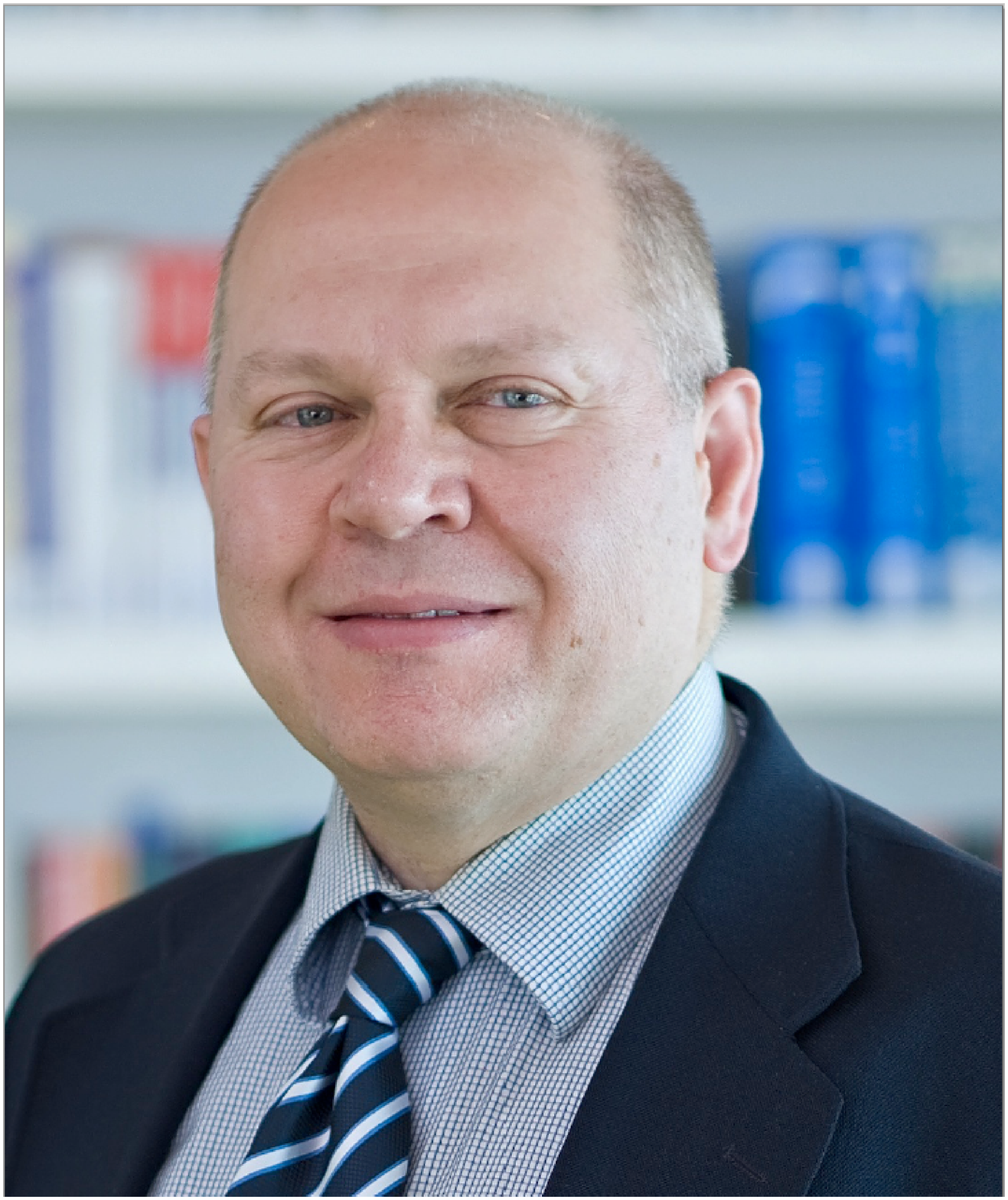}}]{Ibrahim (Abe) M. Elfadel} (SM'-02) received his Ph.D. from Massachusetts Institute of Technology (MIT) in 1993 and is currently Professor and Head of Microsystems Engineering at the Masdar Institute of Science and Technology, Abu Dhabi, UAE. 

He has 15 years of industrial experience with IBM in the research, development and deployment of advanced computer-aided design (CAD) tools and methodologies for deep-submicron, high-performance digital designs. His group's research is concerned with several aspects of energy-efficient digital system design and includes CAD for variation-aware, low-power nano-electronics, power and thermal management of multicore processors, embedded DSP for mmWave wireless systems, modeling and simulation of micro power sources, and 3D integration for energy-efficeint VLSI design. Dr. Elfadel is the Director of the TwinLab/Abu Dhabi Center for 3D IC Design, a joint R \& D  program with the Technical University of Dresden, Germany.  

Dr. Elfadel is the recipient of six Invention Achievement Awards, an Outstanding Technical Achievement Award and a Research Division Award, all from IBM, for his contributions in the area of VLSI CAD.  He is currently serving as an Associate Editor for the IEEE Transactions on Computer-Aided Design for Integrated Circuits and Systems and the IEEE Transactions on Very-Large-Scale Integration.

\end{IEEEbiography}

\begin{IEEEbiography}[{\includegraphics[width=1in,height=1.25in,clip,keepaspectratio]{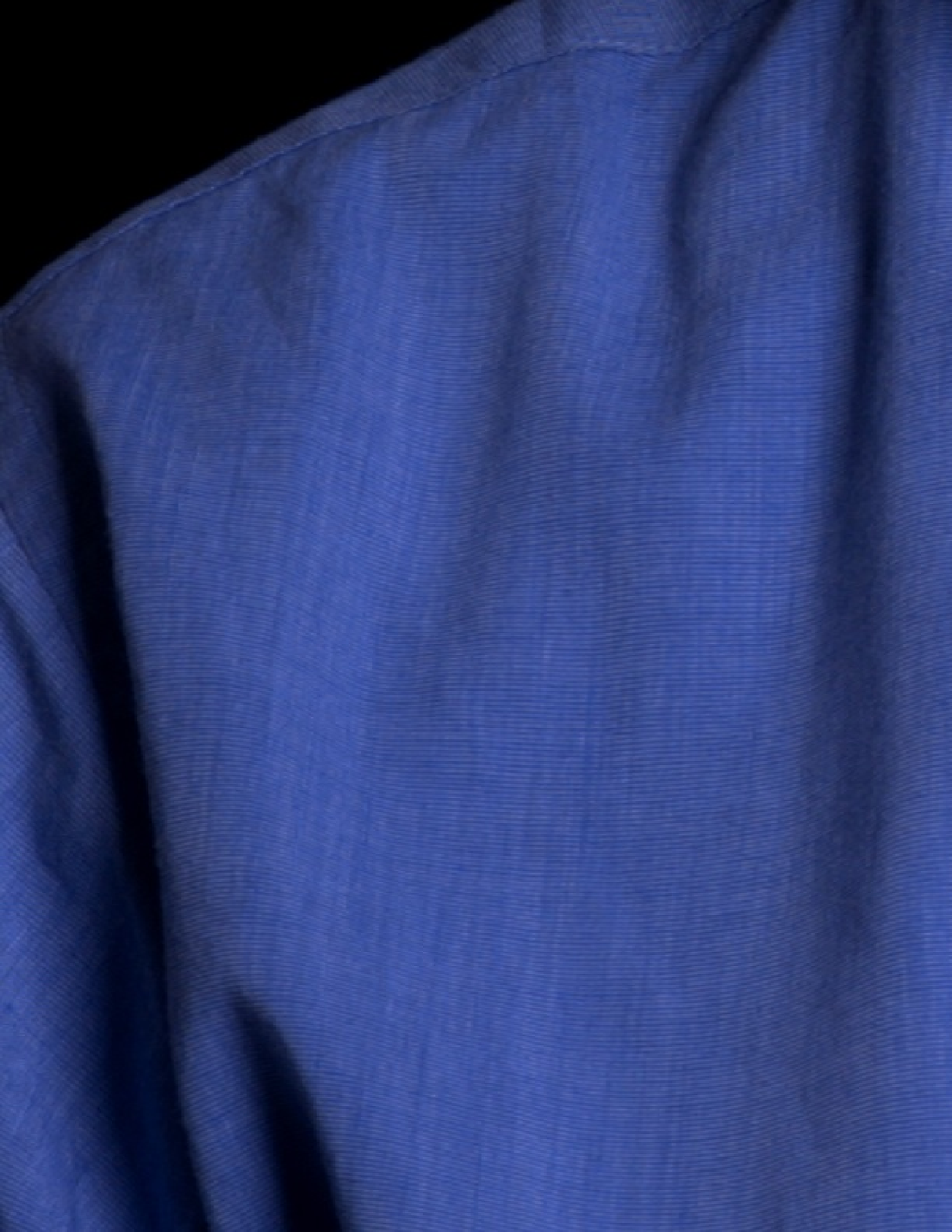}}]{Luca Daniel} (S'98-M'03) received the Laurea degree ({\it summa cum laude}) in electronic engineering from the Universita di Padova, Italy, in 1996, and the Ph.D. degree in electrical engineering from the University of California, Berkeley, in 2003.

He is an Associate Professor in the Electrical Engineering and Computer Science Department of the Massachusetts Institute of Technology (MIT), Cambridge. His research interests include accelerated integral equation solvers and parameterized stable compact dynamical modeling of linear and nonlinear dynamical systems with applications in mixed-signal/RF/mm-wave circuits, power electronics, MEMs, and the human cardiovascular system.

Dr. Daniel received the 1999 IEEE TRANSACTIONS ON POWER ELECTRONICS best paper award, the 2003 ACM Outstanding Ph.D. Dissertation Award in Electronic Design Automation, five best paper awards in international conferences, the 2009 IBM Corporation Faculty Award, and 2010 Early Career Award from the IEEE Council on Electronic Design Automation
\end{IEEEbiography}

\end{document}